\documentclass[aps,prd,twocolumn,nofootinbib,preprintnumbers,floatfix]{revtex4-1}

\usepackage{color}
\definecolor{darkgreen}{rgb}{0,0.5,0}

\newcommand{\DLMass}{10~\textrm{MeV}}
\newcommand{\DLLum}{6 \times 10^{35} \, \textrm{cm}^{-2}\,\textrm{s}^{-1}}
\newcommand{\BackwardsE}{\theta_{\text{cut}}}
\newcommand{\miss}{\text{miss}}
\newcommand{\cut}{\text{cut}}
\newcommand{\MeV}{~\textrm{MeV}}
\newcommand{\GeV}{~\textrm{GeV}}
\newcommand{\Ref}[1]{Ref.~\cite{#1}}
\newcommand{\Refs}[1]{Refs.~\cite{#1}}
\newcommand{\Fig}[1]{Fig.~\ref{#1}}

\newcommand{\Tab}[1]{Table~\ref{#1}}

\newcommand{\Sec}[1]{Sec.~\ref{#1}}

\newcommand{\Eq}[1]{Eq.~(\ref{#1})}

\newcommand{\KE}{\text{KE}}
\newcommand{\alphaEM}{\alpha_{\rm{EM}}}

\usepackage{amsmath}
\usepackage{amssymb}
\usepackage{graphicx}

\begin{document}

\preprint{MIT--CTP 4398}

\title{Searching for an invisible $\boldsymbol{A'}$ vector boson with DarkLight}

\author{Yonatan Kahn}
\email{ykahn@mit.edu}
\author{Jesse Thaler}
\email{jthaler@mit.edu}
\affiliation{Center for Theoretical Physics, Massachusetts Institute of Technology, Cambridge, MA 02139, U.S.A.}

\begin{abstract}
High-luminosity experiments are able to search for new physics at low energies, which could have evaded detection thus far due to very weak couplings to the Standard Model. The DarkLight experiment at Jefferson Lab is designed to search for a new U(1) vector boson $A'$ in the mass range $10-100 \MeV$ through its decay $A' \to e^+ e^-$.  In this paper, we demonstrate that DarkLight is also sensitive to an $A'$ decaying to invisible final states.   We analyze the DarkLight reach for invisible $A'$ bosons assuming a nominal two month running time, including the possibility of augmenting the DarkLight design to include photon detection.  We also propose two new analysis techniques that might prove useful for other high-luminosity searches:  a cut on missing energy to improve the invariant mass resolution, and a cut on the sign of the missing invariant mass-squared to mitigate pileup. We compare the DarkLight reach to existing experimental proposals, including a complementary search using the VEPP-3 positron beam.
\end{abstract}

\maketitle

\section{Introduction}

As the Large Hadron Collider (LHC) continues to expand the energy frontier of particle physics, there are numerous other experiments searching for new physics at the intensity frontier. Accelerators such as the high-current electron beam at the Free Electron Laser (FEL) at Jefferson Lab (JLab) deliver extremely high luminosities, allowing for detailed probes of new physics at MeV-scale energies through precision measurements.

One possibility which has generated much theoretical interest in recent years is a U(1) vector boson $A'$, a spin-1 particle with a mass of $\mathcal{O}(1\MeV-10\GeV)$ which interacts with charged matter via kinetic mixing with the photon.  This idea was first suggested in \Ref{Holdom:1985ag} (see also \Refs{Okun:1982xi,Galison:1983pa}).  Such an $A'$ could mediate annihilation of MeV-scale dark matter $\chi$ into electrons via $\chi \chi \to A' \to e^+ e^-$ \cite{Boehm:2003hm, Pospelov:2007mp} which could explain the excess \cite{Weidenspointner:2006nua} of 511 keV photons from the galactic center \cite{Boehm:2003bt, Huh:2007zw} and a $3\sigma$ anomaly \cite{Abouzaid:2006kk, Dorokhov:2007bd} in the $\pi^0 \to e^+ e^-$ decay rate \cite{Kahn:2007ru}. Alternatively, dark matter could be at the TeV scale, with the 511 keV excess explained with excited dark matter states \cite{Finkbeiner:2007kk, ArkaniHamed:2008qn, Morris:2011dj}.   In such models, pair annihilation $\chi \chi \to A' A'$ followed by the decay $A' \to e^+ e^-$ \cite{Pospelov:2007mp,ArkaniHamed:2008qn} could also explain the high-energy positron excess in the PAMELA \cite{Adriani:2008zr} and FERMI \cite{Abdo:2009zk} data.  In any of these models, indirect constraints from electron and muon anomalous magnetic moments \cite{Fayet:2007ua, Pospelov:2008zw} force the $A'$ to have extremely weak coupling to matter, $\alpha' \approx 10^{-6} \times \alpha_{\rm{EM}}$, allowing the $A'$ to have evaded detection thus far.
 
A light particle like the $A'$ with weak couplings to matter leads to very interesting phenomenology, and several experiments have been designed to search for its distinctive signatures. One possible decay mode is $A' \to e^+ e^-$ (the \emph{visible mode}), inviting a search for low-mass resonances in the $e^+ e^-$ invariant mass spectrum \cite{Reece:2009un, Essig:2009nc, Bjorken:2009mm}.  There are currently several experiments being developed to look specifically for the visible decay mode: the APEX experiment \cite{Essig:2010xa, Abrahamyan:2011gv} and Heavy Photon Search (HPS) \cite{HPS} at the CEBAF facility at JLab, the Hidden Photon Search (HIPS) \cite{Andreas:2010tp} at DESY, the A1 experiment \cite{Merkel:2011ze} at the MAMI facility in Mainz, and the DarkLight\footnote{``Detecting A Resonance Kinematically with eLectrons Incident on a Gaseous Hydrogen Target''} experiment \cite{Freytsis:2009bh,PAC39} at the JLab FEL.   Alternatively, the $A'$ could decay primarily into invisible final states such as neutrinos or dark matter (the \emph{invisible mode}), in which case one must perform a missing energy search.  Such invisible searches were initially proposed in \Ref{Heinemeyer:2007sq}, and there is a recent search proposal using the VEPP-3 positron beam \cite{Wojtsekhowski:2009vz, Wojtsekhowski:2012zq} relevant for both the visible and invisible modes.

In this paper, we show that the DarkLight experiment can be modified in a straightforward way to perform an invisible $A'$ search.  DarkLight, which uses a 100 MeV electron beam incident on a hydrogen gas target to produce an $A'$ through the process $ep \to epA'$, was originally designed to detect an $A' \to e^+ e^-$ resonance in the visible decay mode.  Because DarkLight has full event reconstruction capabilities, it can also detect the single electron and proton which would result from an invisible decay:
\begin{equation}
\label{eq:invprocess}
ep \to ep A', \qquad A' \to \text{inv.}
\end{equation}
Reconstruction of the final-state electron and proton four-vectors could potentially reveal a missing invariant mass peak at the mass of the $A'$, and we will outline a concrete analysis strategy to identify the $A'$.  

The invisible search has extremely challenging backgrounds.  The reconstructed missing invariant mass depends on the momentum of all visible particles, both initial and final states.   The effects of beam energy variation as well as final-state energy and momentum uncertainties lead to a large QED background from $ep \to ep \gamma$ when the photon is unmeasured, a process which would naively have zero missing invariant mass with perfect resolution.\footnote{\label{foot:nonzero}Of course, higher order QED effects generate a small non-zero missing mass from initial- and final-state radiation.}   The high beam luminosity also necessitates a careful treatment of pileup.   To deal with the particular difficulties of a low-energy high-luminosity experiment such as DarkLight, we introduce two new techniques useful for invisible low-mass resonance searches in general:
\begin{itemize}
\item \textbf{Kinematic cuts to improve missing mass resolution:}   In \Sec{sec:InvtMassRes}, we propose an event-selection criterion based on $E_{\miss}/m_{\miss}$, the ratio of missing energy to missing invariant mass, which improves the mass resolution of the experiment. 
\item \textbf{Pileup rejection from negative missing mass-squared:}  In \Sec{sec:Pileup}, we demonstrate that the leading effect of pileup can be eliminated by a cut on the \emph{sign} of the missing invariant mass-squared,\footnote{For clarity, we will often plot missing invariant mass $m$ rather than mass-squared $m^2$, with $m$ defined as $\textrm{sign}(m^2)\times \sqrt{|m^2|}$.} which is strictly negative when an elastic collision ($ep \to ep$) is coincident with any other collision.
\end{itemize}
Finally, to improve rejection of the various QED backgrounds and to prevent bremsstrahlung events from saturating the maximum event reconstruction rate, we will advocate in \Sec{sec:Photon} for augmenting the original DarkLight design with photon detection.

A complementary experiment to detect an $A'$ in either the visible or invisible mode uses the high-intensity positron beam at the VEPP-3 storage ring at the Budker Institute at Novosibirsk incident on a hydrogen target, this time using the atomic electrons for the process $e^+ e^- \to \gamma A'$ \cite{Wojtsekhowski:2012zq}. Here the signal is a ``missing'' photon compared to where it would have been expected from energy-momentum conservation in ordinary pair annihilation $e^+ e^- \to \gamma \gamma$.  This experiment is complementary to DarkLight for several reasons, including the cross section enhancement from the two-body final-state in VEPP-3 compared to three-body state for DarkLight, but the higher $A'$ mass reach in DarkLight compared to VEPP-3.  Although our analysis will focus on DarkLight, we will compare and contrast our results for those of VEPP-3 given in \Ref{Wojtsekhowski:2012zq}.

The remainder of this paper is organized as follows.  In \Sec{sec:DarkLight}, we describe the DarkLight experiment and its operation for both the visible and invisible searches. In \Sec{sec:Signal}, we describe the signal process and illustrate the effects of a cut on $E_{\miss}/m_{\miss}$ on the invariant mass resolution.  We describe the background processes relevant for DarkLight in \Sec{sec:Backgrounds}, including photon backgrounds and the effects of pileup, and we present our analysis strategy in \Sec{sec:Analysis}.  In \Sec{sec:Reach}, we give the anticipated experimental reach for DarkLight with 60 days of running, including a comparison to VEPP-3 and possible constraints from rare kaon decays.  We conclude in \Sec{sec:Conclusions}.

\section{The DarkLight experiment}
\label{sec:DarkLight}

In this section we describe the DarkLight experimental setup, which is optimized to search for visibly-decaying $A'$ bosons, and show how it can be modified in a straightforward way to search for invisible $A'$ decays. 

\subsection{Experiment design and visible search}

\begin{figure*}[tp]
\begin{center}
\includegraphics[width=1.4\columnwidth]{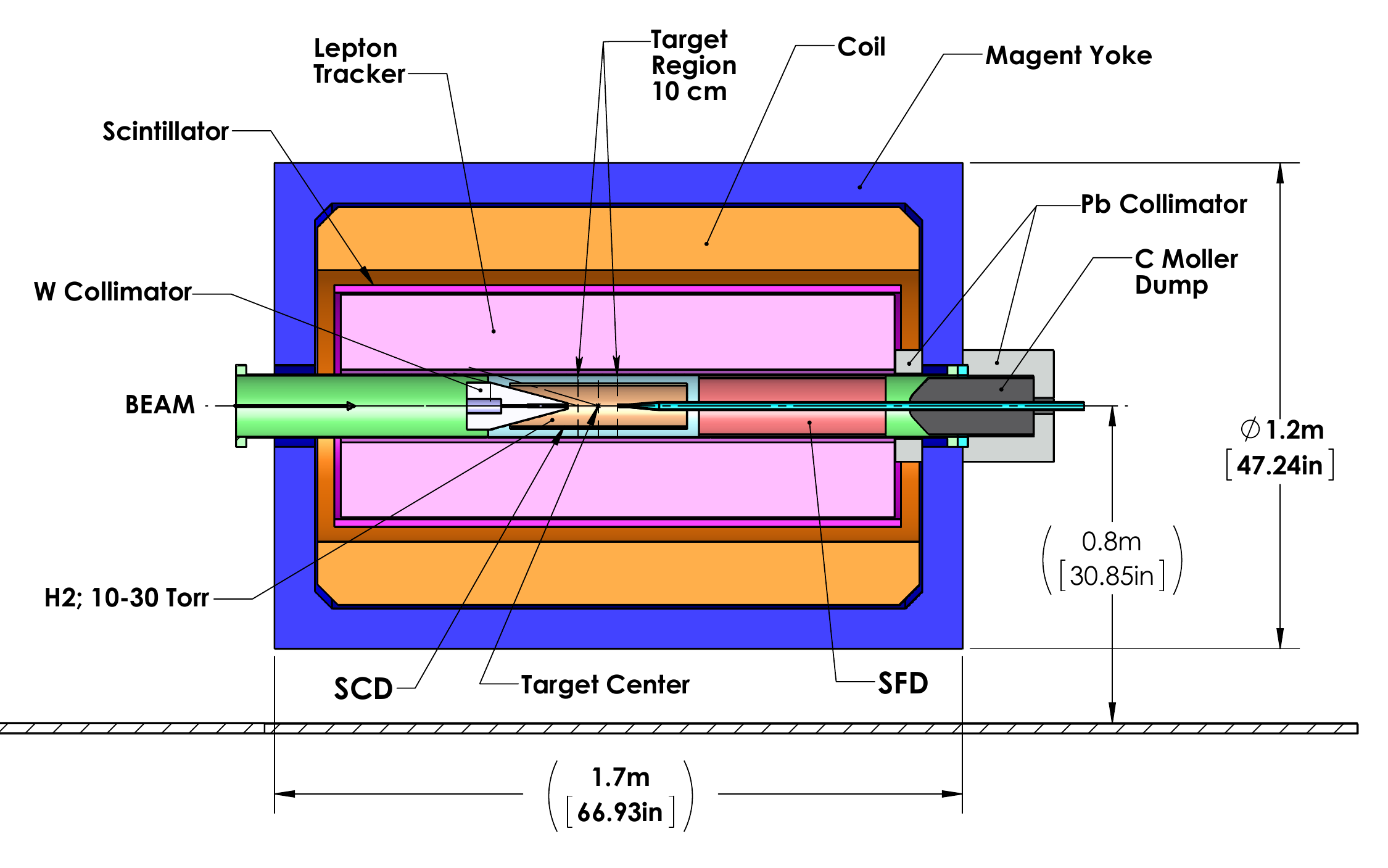}
\caption{Schematic layout of the DarkLight experiment (revised from \Ref{PAC39}). Leptons are tracked in the time projection chamber (TPC, labeled ``lepton tracker'', pink), and forward protons are tracked in the silicon forward detector (SFD, red).}
\label{fig:Detector}
\end{center}
\end{figure*}

\begin{table*}[tdp]
\begin{center}\begin{tabular}{|c||c|c|c|}
\hline \hline
\textbf{Particle} & $\boldsymbol{e^+/e^-}$ &  $\boldsymbol{p}$ & $\boldsymbol{\gamma}$ \\
\hline 
\hline \textbf{Angular acceptance} & $25^\circ-165^\circ$ & $5^\circ-89^\circ$ &  $25^\circ-165^\circ$ \\
\hline \textbf{Angular resolution} & $\sigma_{\theta} = \sigma_{\phi} = 0.002$ rad & $\sigma_{\theta} = \sigma_{\phi} = 0.02$ rad &  N/A\footnote{The photon detection capability is only used as a veto in this analysis, so angular and energy resolutions are not needed.} \\
\hline \textbf{Energy/momentum resolution}& $\sigma_{p_T}/p_T = 2\% \times (p_T/ 100\MeV)$ &  $\sigma_\KE/\KE = 0.1\%$ & N/A$^{\textrm{a}}$ \\
\hline \textbf{Detection threshold}& $p_T > 10 \MeV$ &  $\KE > 1 \MeV$ &  $E > 5 \MeV$ \\
\hline \textbf{Detection efficiency}& 100\%\footnote{The precise electron detection efficiency is not relevant for this analysis, as we will trigger on electrons but veto on protons.}  &  99\% &  95\%\footnote{This is the nominal design efficiency. We will later show how the reach changes quantitatively as a function of photon efficiency.  In reality, photon efficiency varies considerably as a function of photon energy \cite{NarbePrivate}, but taking various benchmark values of flat efficiency will suffice for this analysis.} \\
\hline \hline
\end{tabular}
\caption{Experimental parameters for DarkLight, adapted from \Ref{PAC39}. Angles $\theta$ are measured with respect to the outgoing beam direction, with $\theta = 0^\circ$ forward and $\theta = 180^\circ$ backward.}
\label{tab:DLExp}
\end{center}
\end{table*}

DarkLight consists of a high-intensity 100 MeV electron beam  incident on a gaseous hydrogen target with a nominal luminosity of $\DLLum$ \cite{PAC39}.  This experimental setup was first proposed in \Ref{Heinemeyer:2007sq} and adapted to the $A'$ search in \Ref{Freytsis:2009bh}.  The most recent DarkLight design appears in \Ref{PAC39}.

The target protons provide the recoil momentum for the ``$A'$-strahlung'' process $e^- p \to e^- p A'$.  The $A'$ can then decay into an electron-positron pair, so the final state for the visible search is $e^- p \, e^+ \, e^-$.  A schematic of the detector is shown in \Fig{fig:Detector}.  The electrons and protons are tracked in separate parts of the detector:  electrons with $p_T > 10 \MeV$ are bent in a $0.5~\text{T}$ solenoidal magnetic field and detected in a time projection chamber (TPC) with angular coverage $25^\circ-165^\circ$, while a recoil proton with kinetic energy ($\KE$) greater than 1 MeV is tracked by a silicon forward detector (SFD) with angular coverage $5^\circ-89^\circ$.  Vertexing is accomplished with a silicon central detector (SCD).  The nominal $\phi$ (azimuthal) coverage for both electrons and protons is a full $2\pi$.  The spread in the initial electron beam energy is $\sigma_E/E = 0.1 \%$.   The proton target is at rest, so the only momentum uncertainty is from thermal motion, which is negligible compared to the energy scales of the incoming and outgoing particles.  We summarize the angular coverage, detection thresholds, and experimental resolutions for the various final-state particles in \Tab{tab:DLExp}. Here, detection efficiency will always refer to veto efficiency, and not the efficiency for full object identification.  For example, we will implement a proton veto to mitigate elastic scattering backgrounds, and elastic kinematics dictate precisely where to look for the proton when testing the veto.

A potential $A' \to e^+ e^-$ signal event is tagged by having three charged lepton tracks of the appropriate signs in coincidence with a proton of kinetic energy at least 1 MeV. One then constructs the invariant mass of the two possible electron-positron pairs, looking for a resonance in the $10-90\MeV$ range. The momentum resolution given by the magnetic field is designed so that the invariant mass resolution on $e^+ e^-$ pairs is approximately 1 MeV. A key feature of DarkLight is full four-vector reconstruction of all the final-state particles provided by the silicon detectors and TPC, as well as nearly $4\pi$ angular coverage.  In addition, the 100 MeV beam energy is below the pion threshold, which means that only QED backgrounds are present.

A challenging aspect of this experiment is the huge event rate, combined with the long drift time from the TPC. The event rate is dominated by elastic scattering $ep \to ep$, with an accepted cross section of $1.1 \times 10^{8}$ pb, giving a rate of 65 MHz at nominal luminosity. The TPC drift time is approximately $20~\mu \textrm{s}$, during which there will be on average about 1300 events being tracked. This makes a hierarchical trigger system impractical, and instead one uses a ``free-running'' trigger, where events are pulled from a stream of continuous readout data rather than some trigger object initiating full detector readout.  As a result, one must decide on an event selection criterion which can be applied at data readout time. The presence of three lepton tracks, two negative and one positive, with one of the tracks at an angle $\theta > 60^\circ$ to limit the elastic rate, provides such a criterion for the visible search.  Such a high event rate makes pileup a significant concern, but ``out-of-time'' pileup can be distinguished by timing information from the TPC.  For the purposes of this paper, we model the TPC behavior by assuming that each particle comes with a timestamp accurate to 10 ns, so that the effective window for coincident collisions is 10 ns, rather than the full drift time which is several orders of magnitude longer.

\subsection{Invisible search}
\label{sec:InvisSearch}

\begin{figure}[tc]
\begin{center}
\includegraphics[width=0.8\columnwidth]{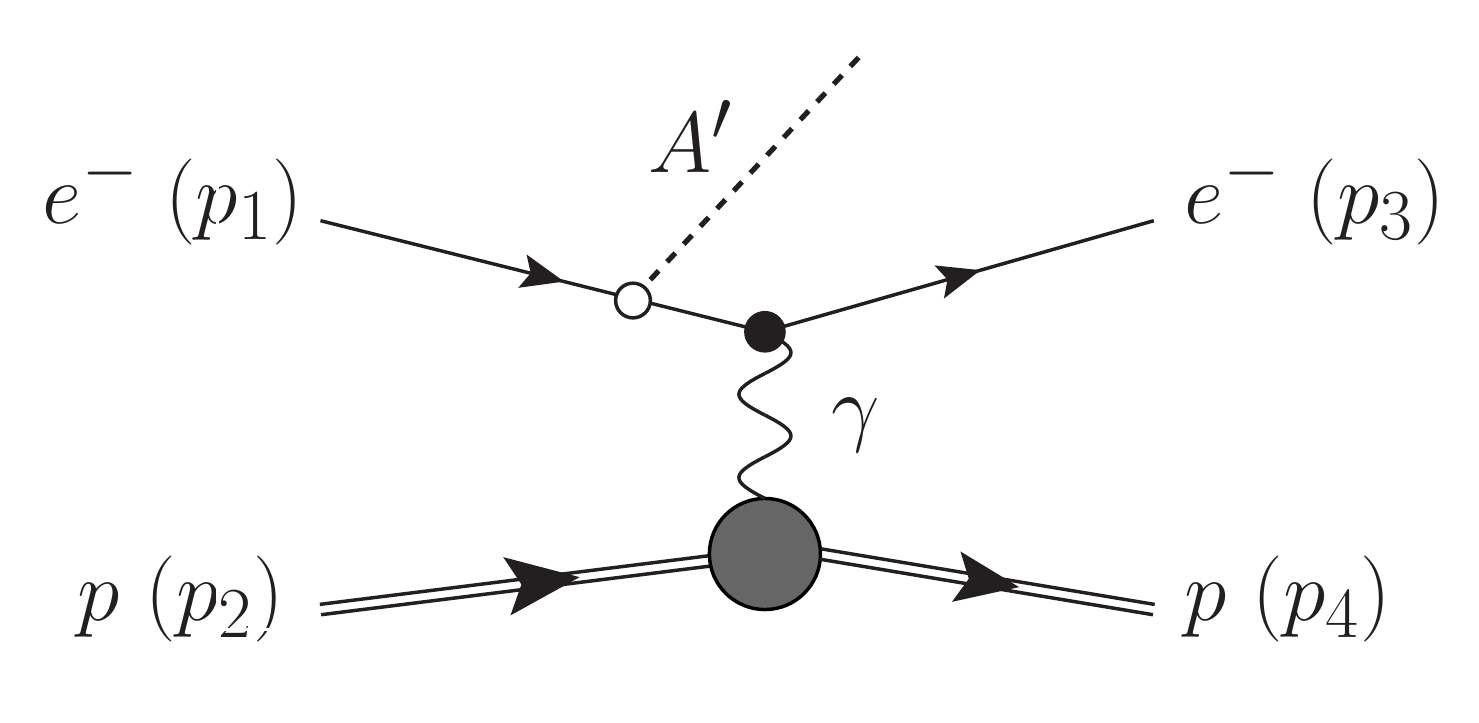}
\caption{Example Feynman diagram for $A'$ production.}
\label{fig:FeynAprimeProd}
\end{center}
\end{figure}

The invisibly-decaying $A'$ mode in \Eq{eq:invprocess} yields a final state consisting of a single electron and a proton.  One can still reconstruct the invariant mass of the missing $A'$ using the final-state four-vectors.  Following \Fig{fig:FeynAprimeProd} and denoting the incoming four-vectors as $p_{e^-} \equiv p_1, p_p \equiv p_2$,  and the outgoing four-vectors as $p'_{e^-} \equiv p_3, p'_p \equiv p_4$, we have
\begin{equation}
m_{\miss}^2 = (p_1 + p_2 - p_3 - p_4)^2.
\end{equation}
The $A'$ invisible search entails looking for a resonance in this missing invariant mass spectrum. 

A principal issue for the invisible search, one which does not pose a problem for the visible search, is that the detector will be swarming with elastic scattering events which also have a single electron and proton in the final state.  Thus, there is no ``smoking gun'' like the positron track of the visible search to provide a readout trigger.  The presence of elastic events which could potentially fake a signal means the effects of pileup must be considered, which we do in \Sec{sec:Pileup}.  A further complication is that the invariant mass we are looking for is no longer defined by a pair of final state particles (i.e.\ the $A' \to e^+ e^-$ resonance), but instead by \emph{four} particles, two of which are incoming and two of which are outgoing. The various kinematic uncertainties associated to all of these four-vectors make a missing invariant mass search extremely challenging, and event selection criteria can somewhat improve the invariant mass resolution, as we discuss in \Sec{sec:InvtMassRes}.

Finally, given the large luminosity of DarkLight, one must account for practical limits on event reconstruction and data acquisition rates \cite{FisherPrivate}.  For the analysis strategy proposed in \Sec{sec:Analysis}, we will assume that the free-running trigger can make one-bit object identification decisions at an arbitrarily fast rate.  We then impose a 50 kHz cap on the rate for full event reconstruction, which also sets the rate at which invariant mass calculations can be made.  Finally, we impose a 300 Hz cap on final event selection and data storage, though the precise amount of data that can be written to tape may ultimately be higher.

\subsection{Photon detection}
\label{sec:Photon}

If the DarkLight apparatus is only sensitive to charged particles, as is necessary for the visible search, then QED processes such as $ep \to ep \gamma \gamma$ pose an irreducible background\footnote{Our terminology here is slightly nonstandard: ``irreducible'' will refer to any background process whose final state is indistinguishable from the signal given the experimental detection capabilities.  An irreducible background for the invisible search is any process where the detector only registers one electron and one proton.} for the invisible search, since the diphoton invariant mass spectrum is broad and covers the same mass range as the $A'$ search.  Even the single-photon process $ep \to ep \gamma$, which is in principle reducible since the photon has zero invariant mass, becomes an irreducible background due to errors on the reconstructed invariant mass, as we will show in \Sec{sec:PhotonBackgrounds}. Thus, it is highly advantageous to have some kind of photon detection capability. 

The most recent iteration of the DarkLight design includes a simple calorimeter surrounding the TPC at a distance of 30 cm from the beamline, consisting of three layers of lead interleaved with scintillator material. The angular coverage of such a photon detector would nominally be the same as the electron coverage, $25^\circ-165^\circ$, but could in principle be extended somewhat in the forward direction.  The backwards angle of $165^\circ$ is set by the geometry of the beam collimator, shown in white in \Fig{fig:Detector}, and cannot be extended further with the current detector design.\footnote{The scintillator shown in \Fig{fig:Detector} only extends to  $\simeq 155^\circ$, so to get up to $165^\circ$ coverage, we assume the existence of a backwards photon endcap, or that the geometry will be adjusted to extend the backwards angular coverage.}  Note that events with photons that miss the detector entirely will still pose an irreducible background; we will make some comments on how the reach improves with increased photon angular coverage in \Sec{sec:Improve}.

This three-layer scintillator design is capable of detecting photons of energies $E_\gamma > 5 \MeV$, with better than $50\%$ efficiency for $E_\gamma > 30 \MeV$ and approximately $95\%$ detection efficiency for photons sufficiently energetic to give a large missing invariant mass \cite{NarbePrivate}. The efficiency can be improved by adding more layers of scintillator, and to be optimistic we will take $95\%$ as a benchmark value. In our analysis, we define a photon as having energy greater than 5 MeV, and treat $ep\gamma$ events with soft photons ($E_{\gamma} < 5 \MeV$) as contributing to the next-to-leading order (NLO) correction to the elastic process.\footnote{In this paper, we restrict to leading-order (i.e.~tree-level) calculations for background cross sections; the cut on photon energy at 5 MeV keeps the QED cross sections safe from large logarithms and ensures that the NLO corrections are manageable.  Restricting to leading-order calculations is justified, since NLO corrections are not expected to generate a kinematic feature in the $m_\miss$ distribution, though the overall NLO background rate and shape would of course be different.}  We will often call events with soft photons ``quasi-elastic'', though there is no real way to discriminate between the quasi-elastic and elastic processes.  There is a potential issue of electrons faking photons, though an electron must have a $p_T$ of at least 45 MeV to even reach the scintillator given the large magnetic field, and the chance of missing a track of this momentum is quite small.  Moreover, the photon detector will only be used as a veto in our study, and fakes correspond to some small loss of signal efficiency without introducing any new backgrounds.

\section{Signal}
\label{sec:Signal}

\subsection{Rates and kinematics}

\begin{figure}[t]
\begin{center}
\includegraphics[width=0.8\columnwidth]{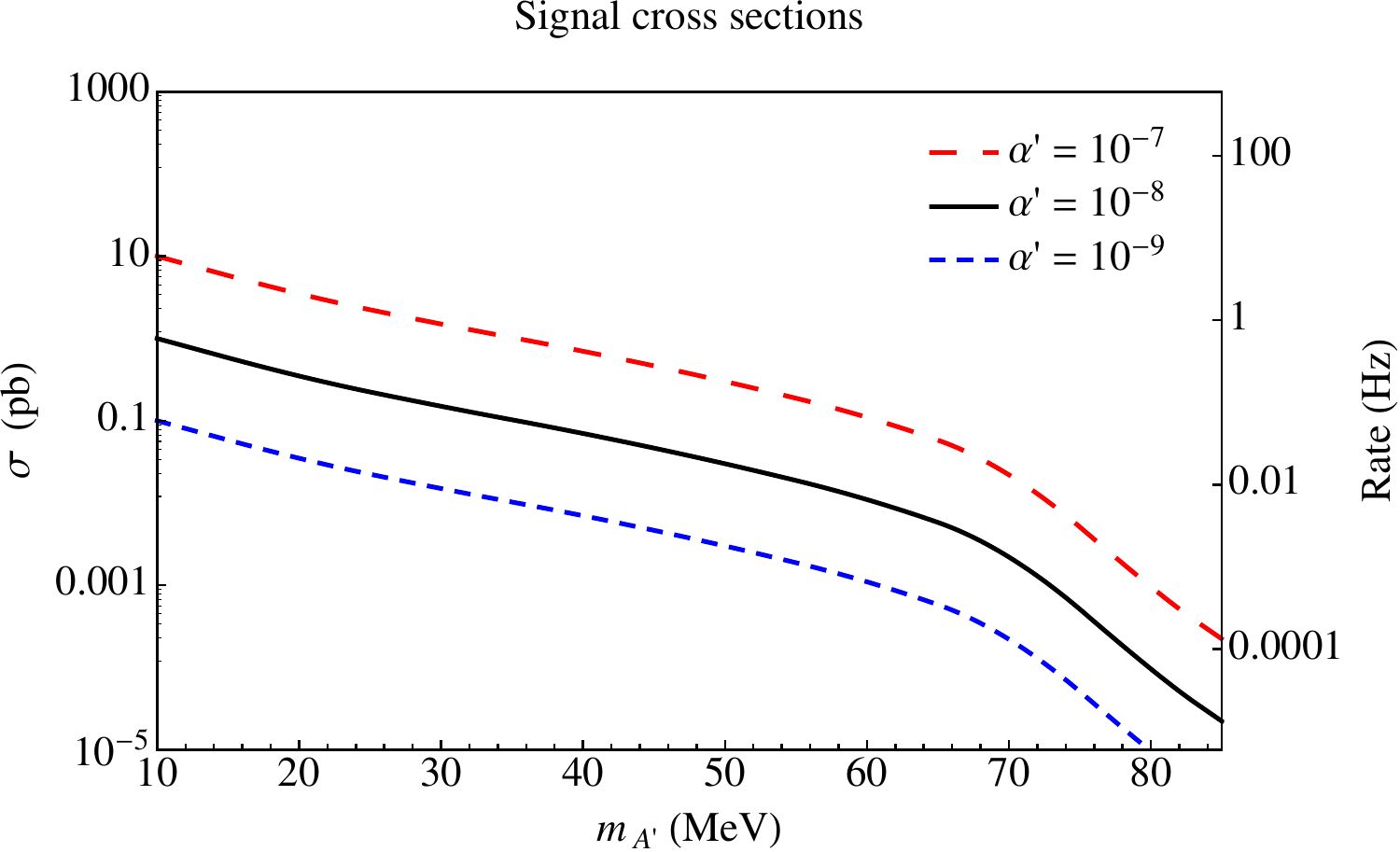} 
\caption{Detector-accepted cross sections for the signal process $ep \to ep A'$ at various $A'$ masses.  The rates on the right-side axis assume the nominal DarkLight luminosity of $\DLLum$.}
\label{fig:SignalRates}
\end{center}
\end{figure}

\begin{figure*}[t]
\begin{center}
\includegraphics[width=0.8\columnwidth]{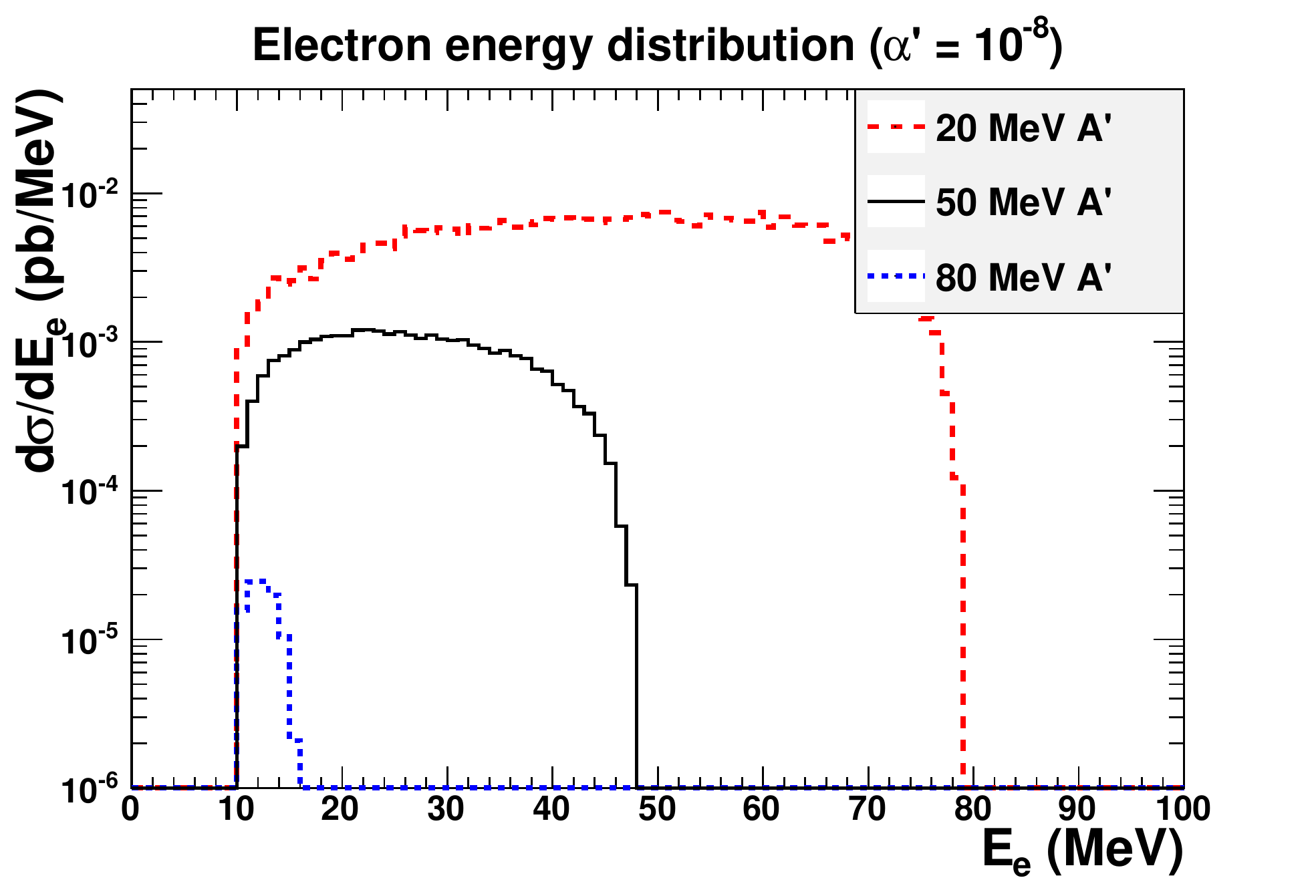} \qquad
\includegraphics[width=0.8\columnwidth]{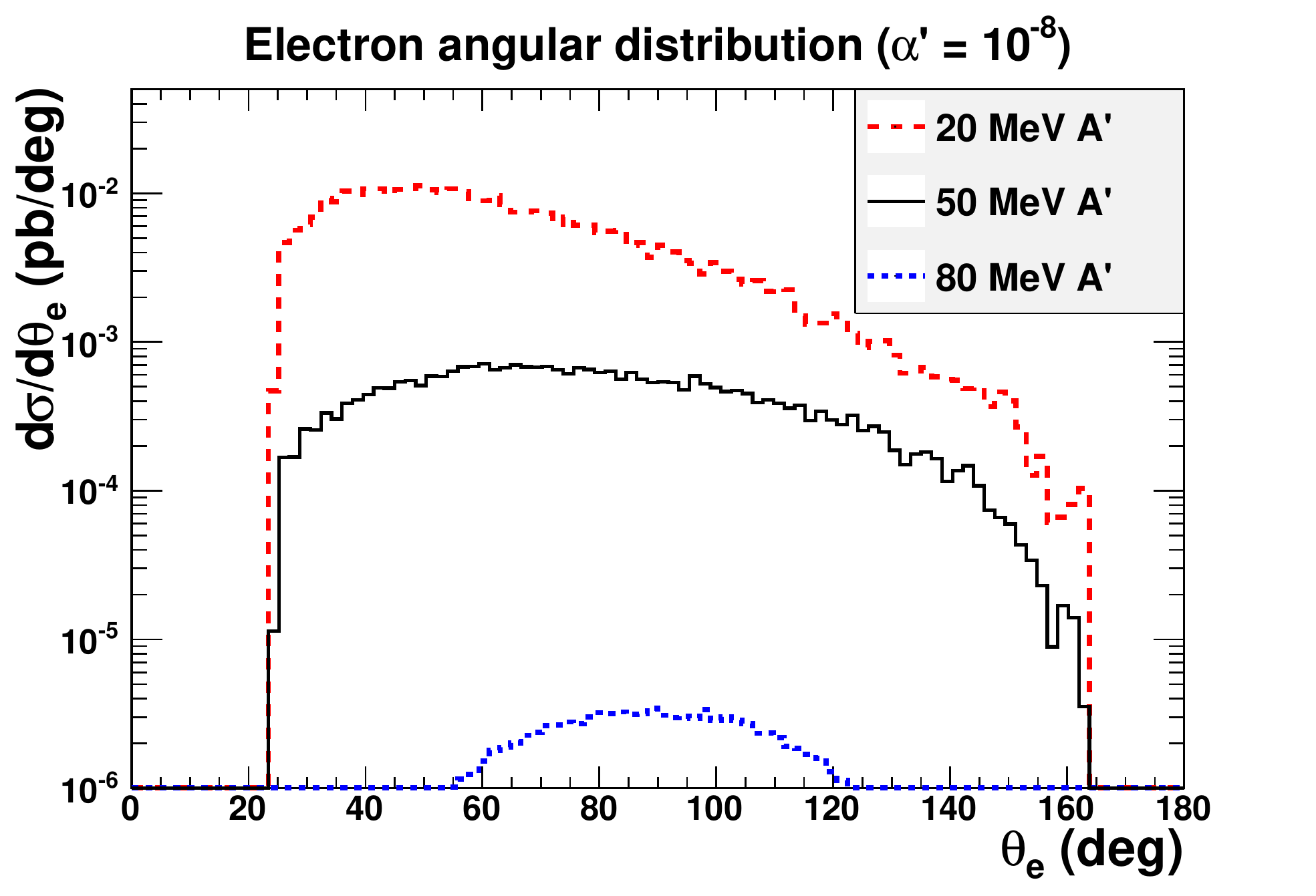} \linebreak
\includegraphics[width=0.8\columnwidth]{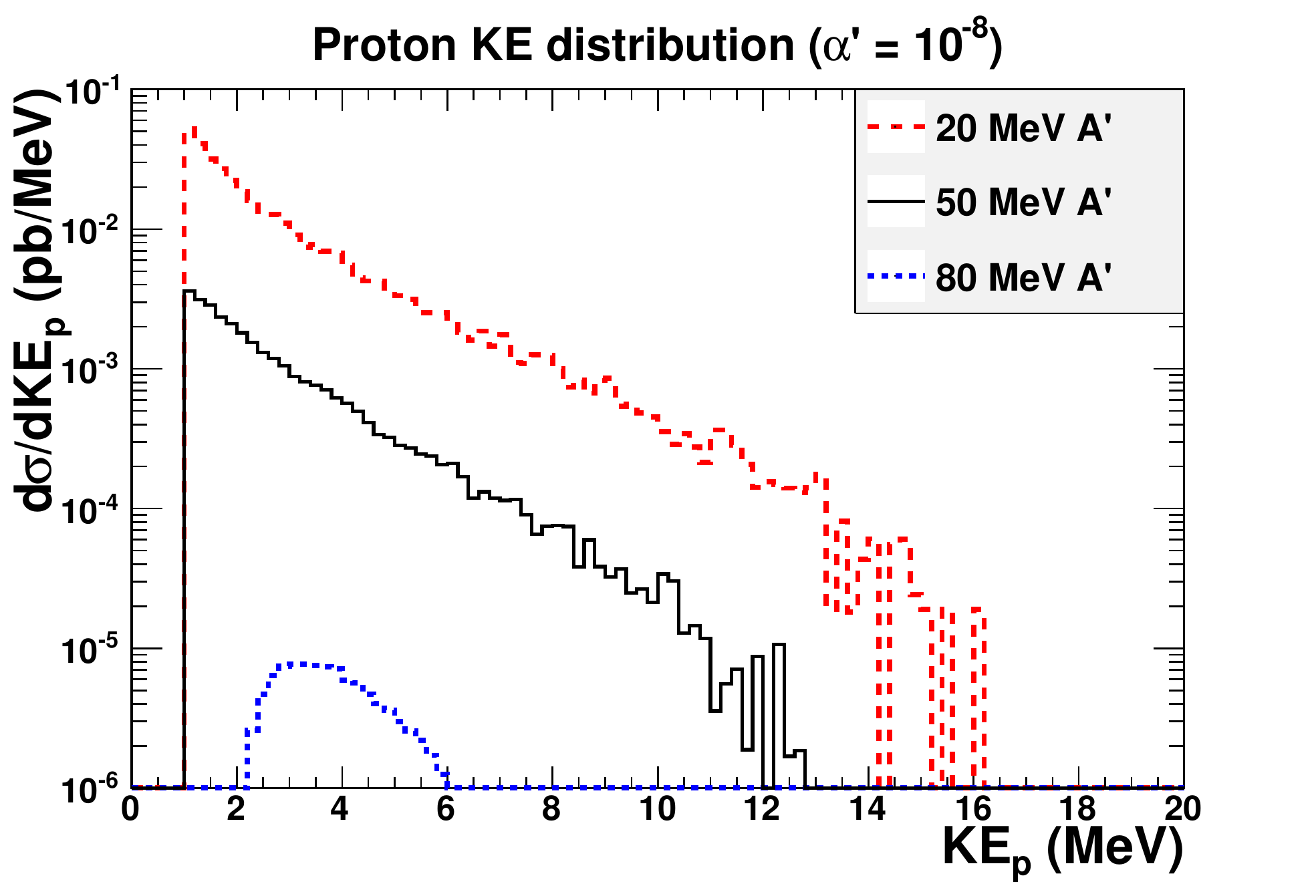} \qquad
\includegraphics[width=0.8\columnwidth]{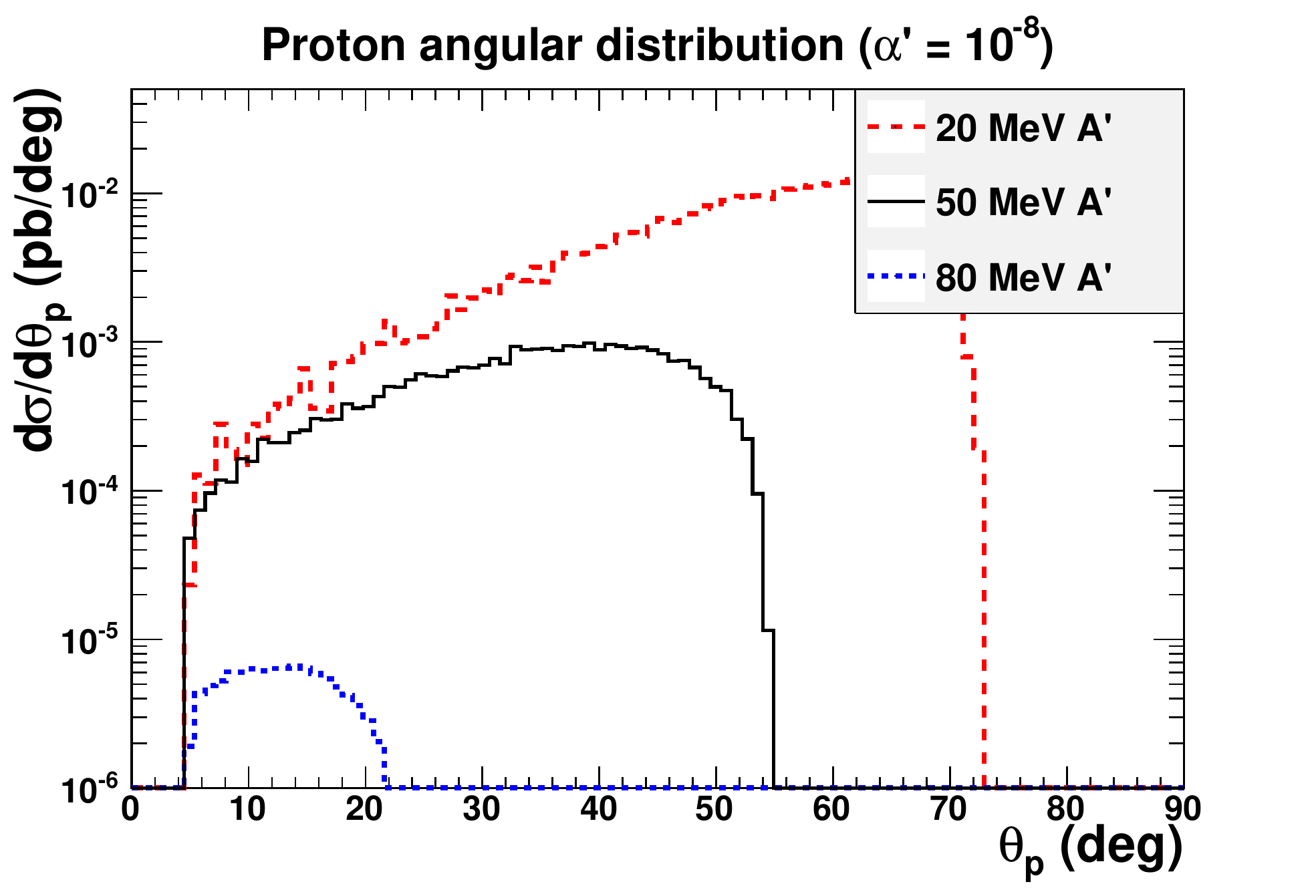} 
\caption{Final-state energy (left) and angular (right) distributions for electron (top) and proton (bottom) for the $epA'$ signal process with DarkLight detector acceptances.}
\label{fig:EPDist}
\end{center}
\end{figure*}

The signal process for the DarkLight invisible search is $ep \to ep A'$, with an example Feynman diagram shown in \Fig{fig:FeynAprimeProd}.  For this analysis we assume a 100\% branching ratio of the $A'$ to invisible decay products; one can easily relax this restriction by adding appropriate factors of $\text{Br}(A' \to \text{inv.})$. We assume the $A'$ couplings to matter originate from kinetic mixing with the photon \cite{Holdom:1985ag}, such that the $A'$ couples universally to charged matter with strength $\alpha'$.   In particular, the $A'$ couples to the proton as well, giving additional diagrams where the proton radiates an $A'$.\footnote{We ignore the proton form factors in this analysis, the effects of which only modulate the signal and background cross sections by percent-level corrections at these low beam energies without introducing any sharp kinematic features. It is consistent to neglect form factors as we are also neglecting NLO corrections to the background cross sections.}  One might expect such diagrams to be suppressed by the proton mass as in ordinary bremsstrahlung, but they in fact generate percent-level corrections for larger masses $m_{A'}$.

The cross sections and associated rates given the angular and energy/momentum acceptances of the DarkLight detector have been calculated using the software packages MadGraph 4.4.44 \cite{Alwall:2007st} and CalcHEP 2.3 \cite{Pukhov:2004ca}.  The total signal cross sections including detector acceptance are shown as a function of $m_{A'}$ in \Fig{fig:SignalRates}.  We have verified that the results from both programs agree to within Monte Carlo uncertainties for all processes of interest.  For the remainder of this paper, we will take a fiducial value of $\alpha' = 10^{-8}$ for any signal cross sections or distributions, except in the analysis example of \Sec{sec:ExampleRates} and when we calculate the $\alpha'$ reach in \Sec{sec:DLReach}.

In \Fig{fig:EPDist} we show the kinematic distributions for the visible final-state particles in the signal process. Unlike elastic scattering or ordinary bremsstrahlung, where the cross section drops sharply as a function of electron angle, the electron kinematic distributions in $A'$ production are much flatter. This suggests an analysis strategy where we focus on wide-angle electrons for potential signal events, to cut down on potential background from mismeasured or misidentified QED events. 

Indeed, a backwards electron cut $\theta_e > \BackwardsE$ is necessary in order to reduce the overall event reconstruction rate to a manageable level, because of the dominant $ep \to ep \gamma$ background.   As we will see in \Sec{sec:Elastic}, the choice of $\BackwardsE = 103^\circ$ saturates the maximum event reconstruction rate of 50 kHz with a photon detection efficiency of 95\%.  We will also see that the minimum required value of $\BackwardsE$ changes as a function of photon efficiency; for illustration purposes we will assume the value corresponding to 95\% efficiency, and relax this restriction in our reach plots in \Sec{sec:Reach}.

\subsection{Improving invariant mass resolution}
\label{sec:InvtMassRes}

A key criterion for the invisible search is sharp missing invariant mass resolution, and we can define our event selection to only maintain signal events that have a high likelihood of successful reconstruction.  Given errors $\sigma_E$ ($\sigma_p$) on the reconstructed energy $E$ (momentum $p$), the reconstructed mass $m = \sqrt{E^2 - p^2}$ will have errors which go like
\begin{equation}
\label{SigmaM}
\sigma^2_m  = \left ( \frac{E}{m} \right)^2 \sigma_E^2 \oplus \left ( \frac{p}{m} \right)^2 \sigma_p^2.
\end{equation}
The errors on the right-hand side are proportional to the ratios $E/m$ and $p/m$, so we define
\begin{equation}
E_{\miss}/m_{\miss} \equiv 1 + \Delta,
\end{equation} 
the ratio of missing energy to missing invariant mass, as a figure of merit.\footnote{We use $E/m$ rather than $p/m$ for convenience, but clearly large $p/m$ implies large $E/m$ and vice versa.} Since $\Delta$ determines the effects of mass mis-measurement, fixing a cut value $\Delta_{\cut}$ and requiring $\Delta < \Delta_{\cut}$ improves the invariant mass resolution of the experiment for fixed $\sigma_E$ and $\sigma_p$.

\begin{figure}[t]
\begin{center}
\includegraphics[width=0.8\columnwidth]{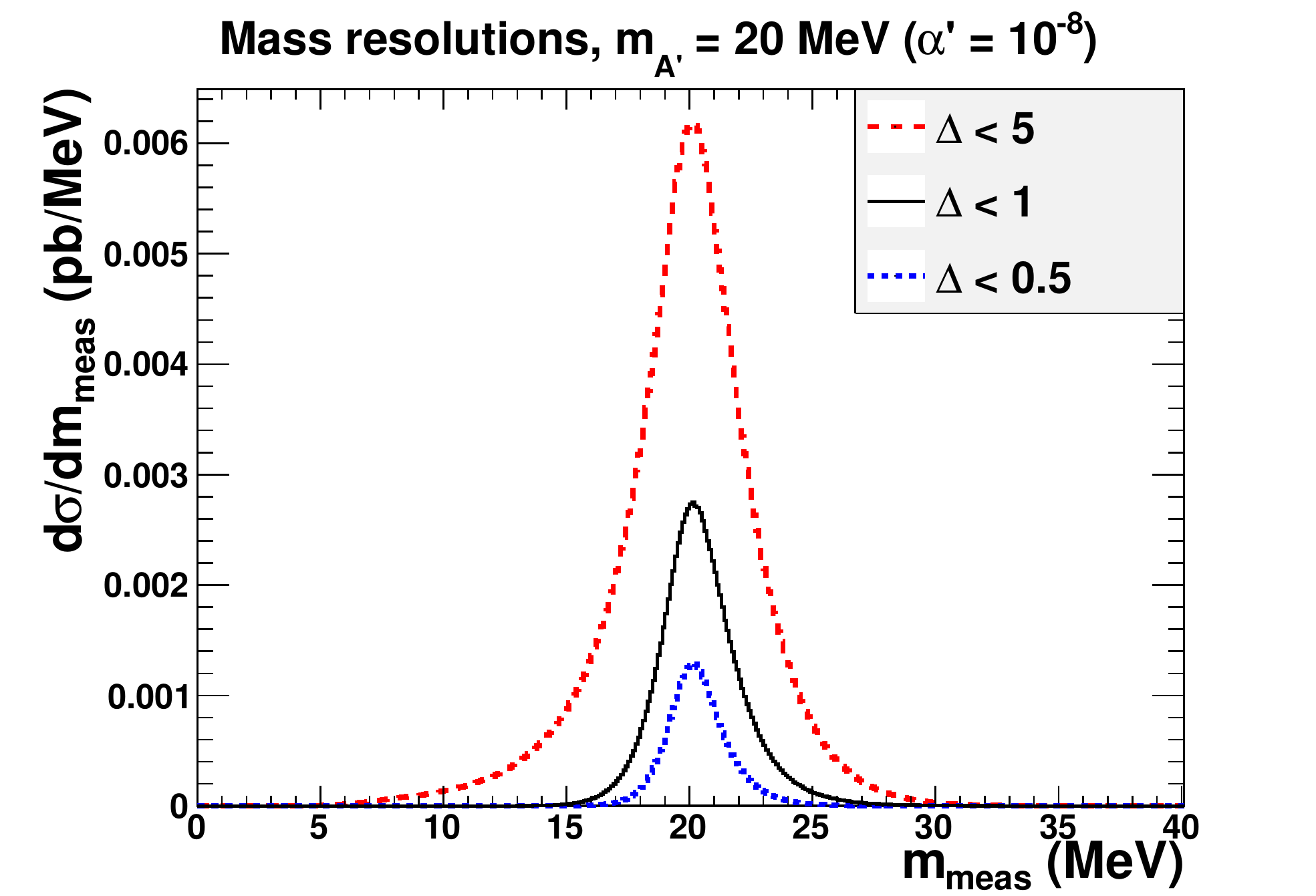}
\caption{Mass resolutions for $m_{A'} = $ 20 MeV with various $\Delta$ cuts. The largest value of $\Delta$ shown is the largest kinematically allowed value, such that no part of the cross section is cut.}
\label{fig:AprimeMassRes}
\end{center}
\end{figure}

\begin{table}[t]
\begin{center}\begin{tabular}{|c||c|c|c|c|c|c|c|c|} \hline \hline
$\mathbf{m_{A'}}$ \textbf{(MeV)} & \textbf{10} & \textbf{20} & \textbf{30} & \textbf{40}& \textbf{50} & \textbf{60} & \textbf{70} & \textbf{80} \\
\hline
\hline $\mathbf{\Delta_{\cut} = 10}$ & 4.0 & 3.7 & 2.6 & 1.8 & 1.3 & 0.92 & 0.66 & 0.35 \\
\hline $\mathbf{\Delta_{\cut} =  1}$ & 3.0 & 2.5 & 2.1 & 1.6 & 1.3 & 0.92 & 0.66 &  0.35 \\
\hline $\mathbf{\Delta_{\cut} =  0.5}$ & 2.1 & 1.8 & 1.5 &  1.4 & 1.2 & 0.92 & 0.66 & 0.35 \\
\hline \hline
 \end{tabular} \caption{Mass resolutions in MeV for various values of $\Delta_{\cut}$.  These resolutions are estimated from the standard deviation of the reconstructed invariant mass distribution.}
\label{tab:MassRes}
\end{center}
\end{table}

In \Fig{fig:AprimeMassRes} we show how the shape of the invariant mass distribution changes as a function of $\Delta_{\cut}$ for the case $m_{A'}  = 20~\MeV$, applying Gaussian smearing via the experimental resolutions in \Tab{tab:DLExp}.  For other values of $m_{A'}$, the mass resolutions for various $\Delta_{\cut}$ are given in \Tab{tab:MassRes}.  Note that by kinematic constraints, the $A'$ can never have more than 100 MeV of lab frame energy, so there is always a value of $\Delta_{\cut}$ above which the signal acceptance is unaffected.  Similarly, since the signal region is restricted to $m_{A'} > 10 \MeV$ by constraints on the photon measurement (see \Sec{sec:PhotonBackgrounds}), a global cut $\Delta < 10$ keeps all of the signal cross section through the accessible mass range.  

\begin{figure}[tp]
\begin{center}
\includegraphics[width=0.8\columnwidth]{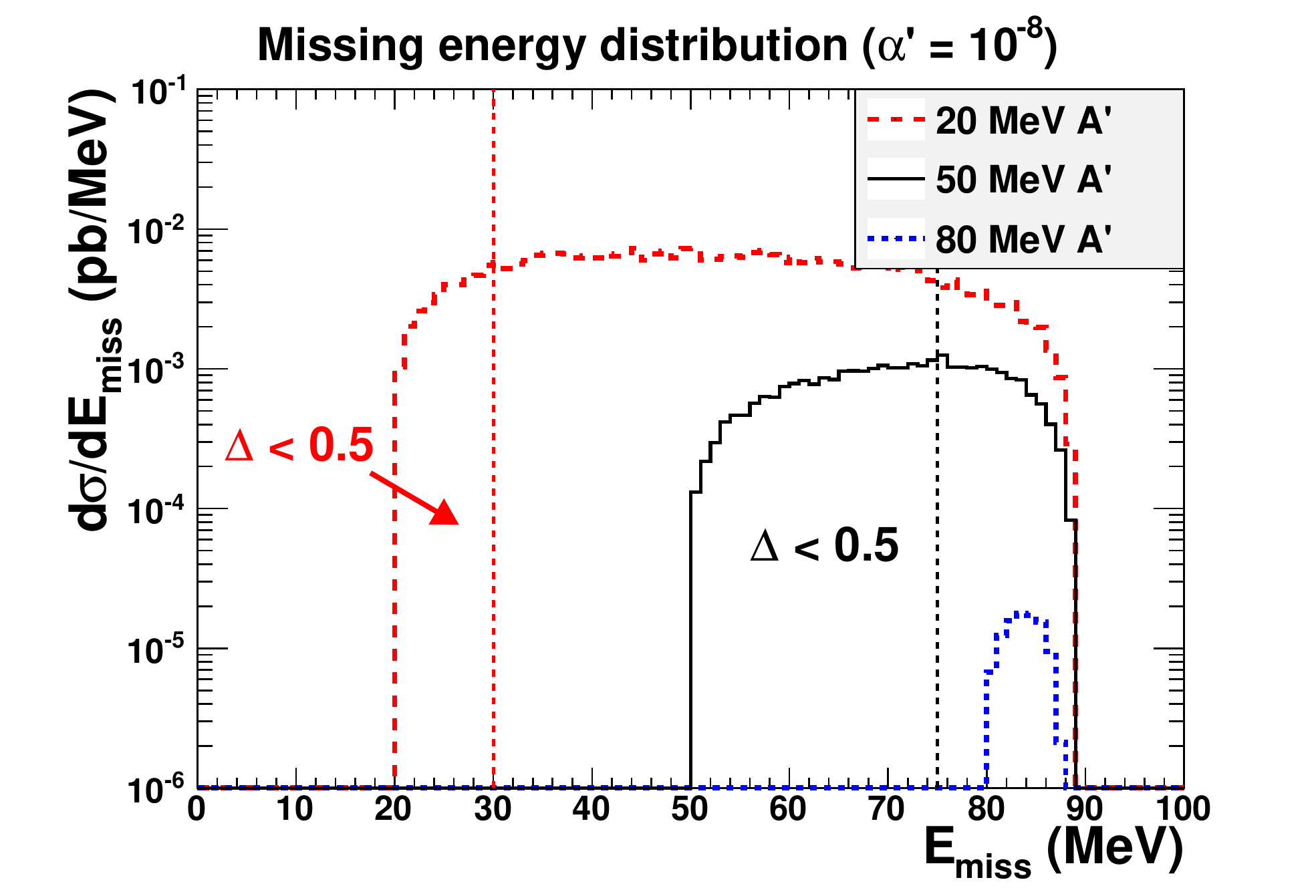}
\caption{Missing energy distributions for the signal process $epA'$.  The portions of the signal kept after a cut of $\Delta < 0.5$ are illustrated by dashed vertical lines. Note that for this particular cut, all of the 80 MeV signal is preserved.}
\label{fig:AprimeDist}
\end{center}
\end{figure}

\begin{figure}[tp]
\begin{center}
\includegraphics[width=0.8\columnwidth]{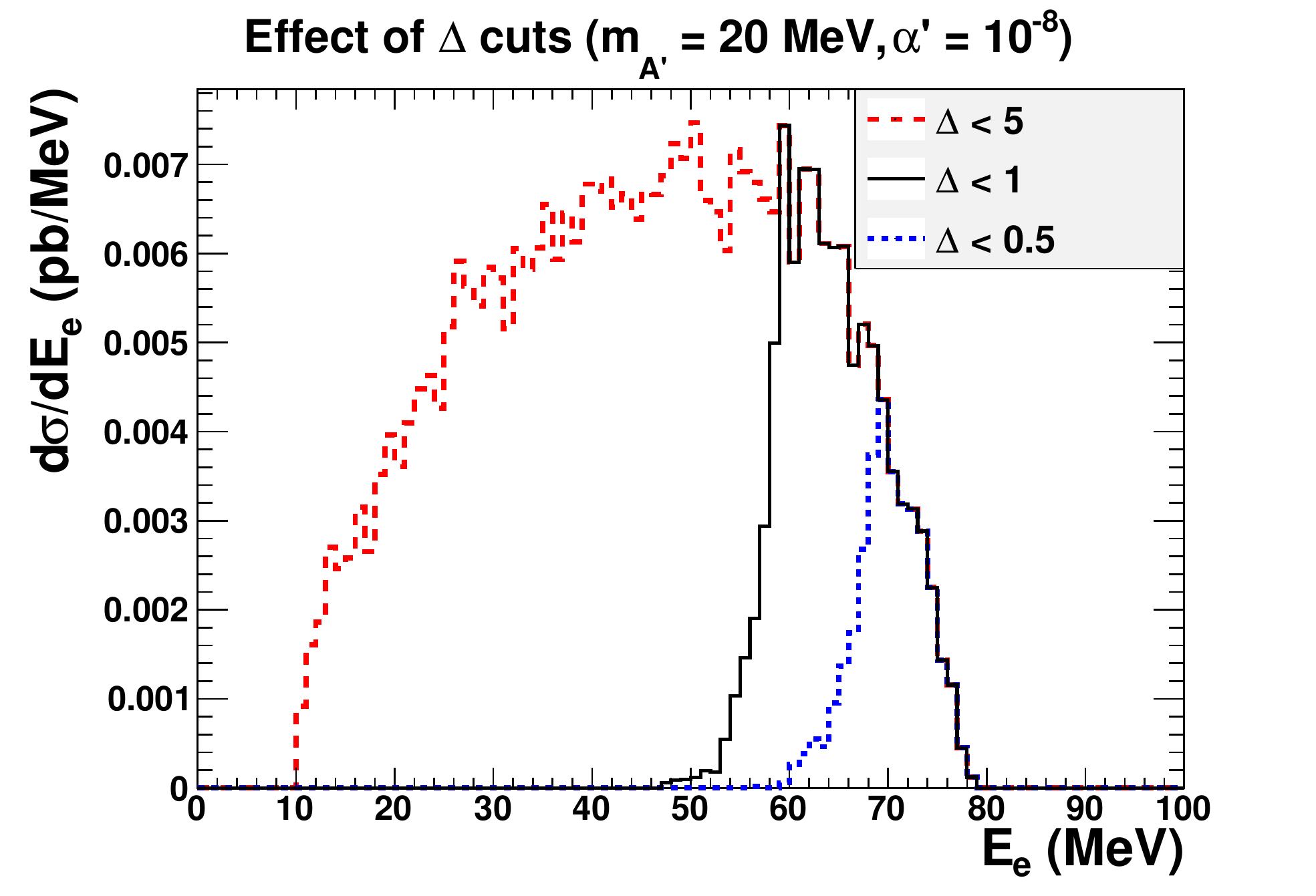} 
\caption{Final-state electron energy distributions for $m_{A'} = 20 \MeV$ with various $\Delta$ cuts.}
\label{fig:EDistDeltaCuts}
\end{center}
\end{figure}

Clearly, stronger cuts on $\Delta$ reduce the signal cross section: in \Fig{fig:AprimeDist}, we see that a cut at $\Delta = 0.5$ (so $E_{A'}/m_{A'} < 1.5$) keeps only about 10\% of the signal for $m_{A'} = 20 \MeV$, but keeps the majority of the signal for $m_{A'} = 50 \MeV$.  In \Sec{sec:DLReach}, we will see that the best reach is obtained for no $\Delta$ cut at all, but we expect that a mild $\Delta < 1$ cut would help build confidence in any purported $A'$ signal.   In \Fig{fig:EDistDeltaCuts}, we show the effect of various $\Delta$ cuts on the electron energy distribution for $m_{A'} = 20 \MeV$; as expected, requiring the $A'$ to be produced nearly at rest forces the electron to come out with more energy.

\section{Backgrounds}
\label{sec:Backgrounds}

We now consider several possible sources of background for the invisible search.  We first consider various types of singles events (two-photon, one-photon, and elastic), before moving on to a careful treatment of pileup. We then consider other potential sources of QED and electroweak backgrounds, and argue that they are negligible compared to the photon and pileup backgrounds. As mentioned in the introduction, a beam energy of 100 MeV puts DarkLight below pion threshold, so there are no QCD backgrounds to contend with.

\subsection{Photon backgrounds}
The principal QED backgrounds involve one or more photons in the final state. Cross sections for photon production have a collinear singularity in the limit where the electron is massless, but in our analysis the beam energy is low enough that we must keep the electron mass, and the singularity is cut off. There is also a soft singularity as the photon energy goes to zero, but we account for this by treating events with $E_{\gamma} < 5 \MeV$ as NLO corrections to lower-order processes, as mentioned in \Sec{sec:Photon}.
\label{sec:PhotonBackgrounds}

\subsubsection{Two-photon}

\begin{figure}[tp]
\begin{center}
\includegraphics[width=0.8\columnwidth]{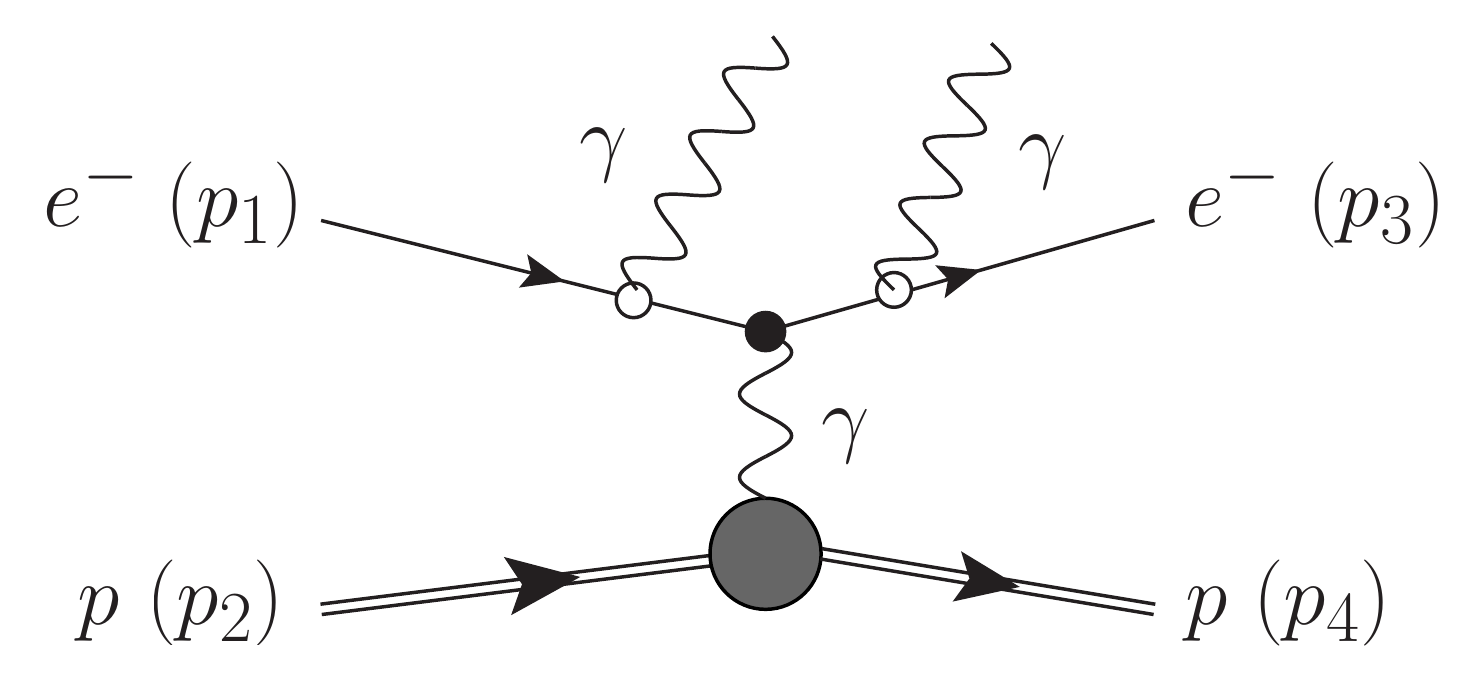}
\caption{Example Feynman diagram contributing to the background $ep \to ep\gamma\gamma$.}
\label{fig:Feynepgg}
\end{center}
\end{figure}

\begin{figure}[tp]
\begin{center}
\includegraphics[width=0.8\columnwidth]{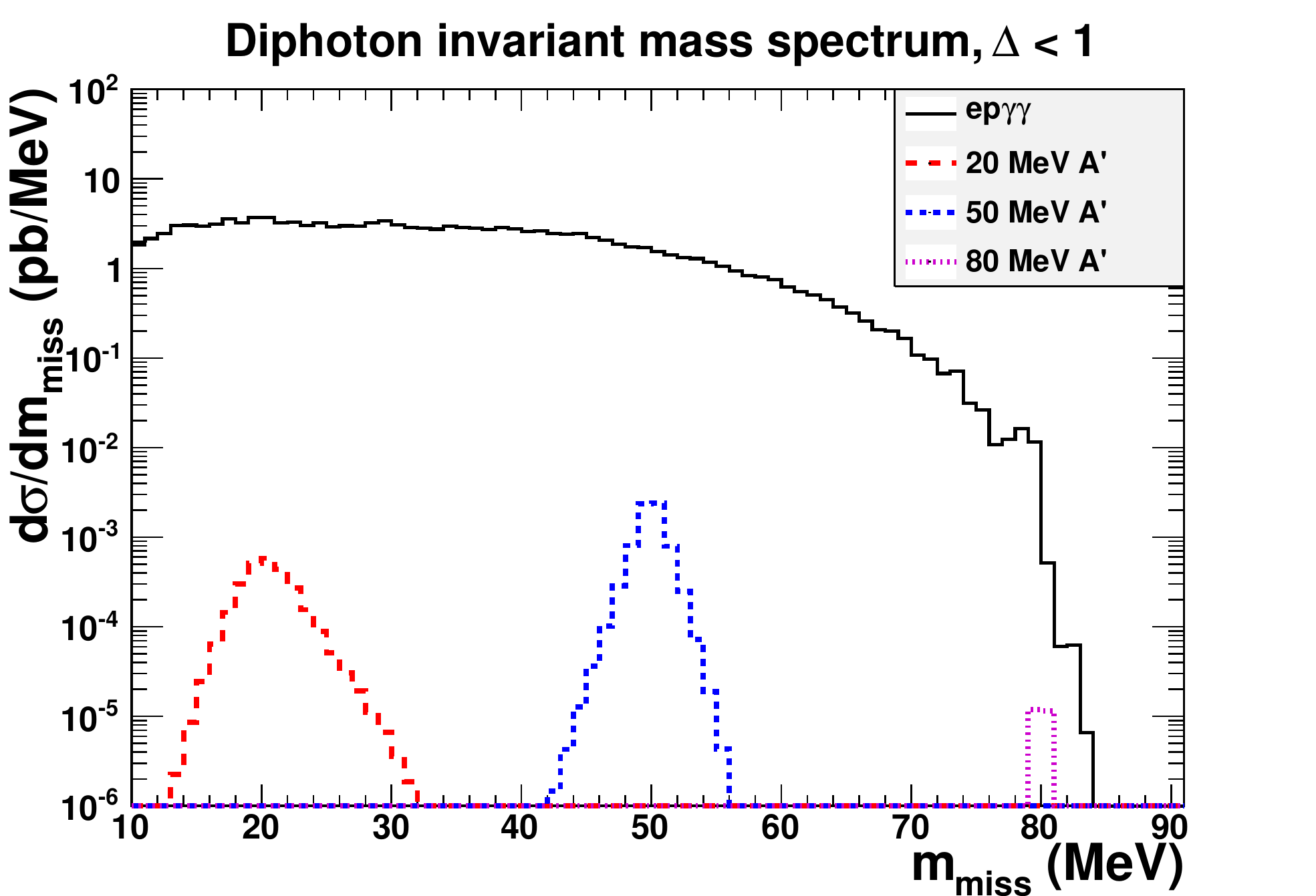}
\caption{Diphoton invariant mass spectrum for $\Delta < 1$ and experimental acceptances. Signal distributions for three values of $m_{A'}$ with $\alpha' = 10^{-8}$ and $\Delta < 1$ are shown for comparison.  The structure in the background at $m_{\miss} = 80\MeV$ is an artifact of low Monte Carlo statistics.}
\label{fig:DiphotonMass}
\end{center}
\end{figure}

\begin{figure*}[tp]
\begin{center}
\includegraphics[width=0.8\columnwidth]{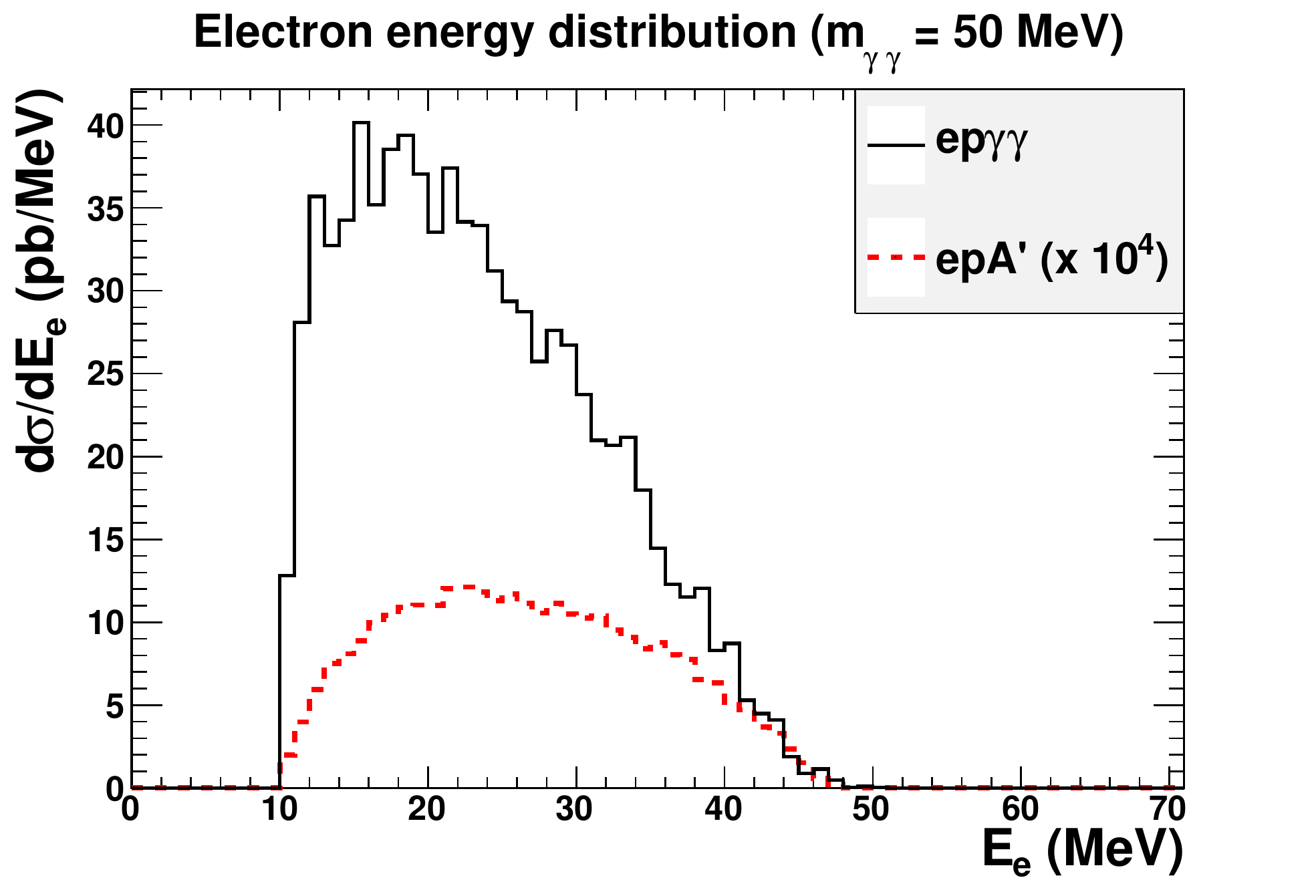} \qquad
\includegraphics[width=0.8\columnwidth]{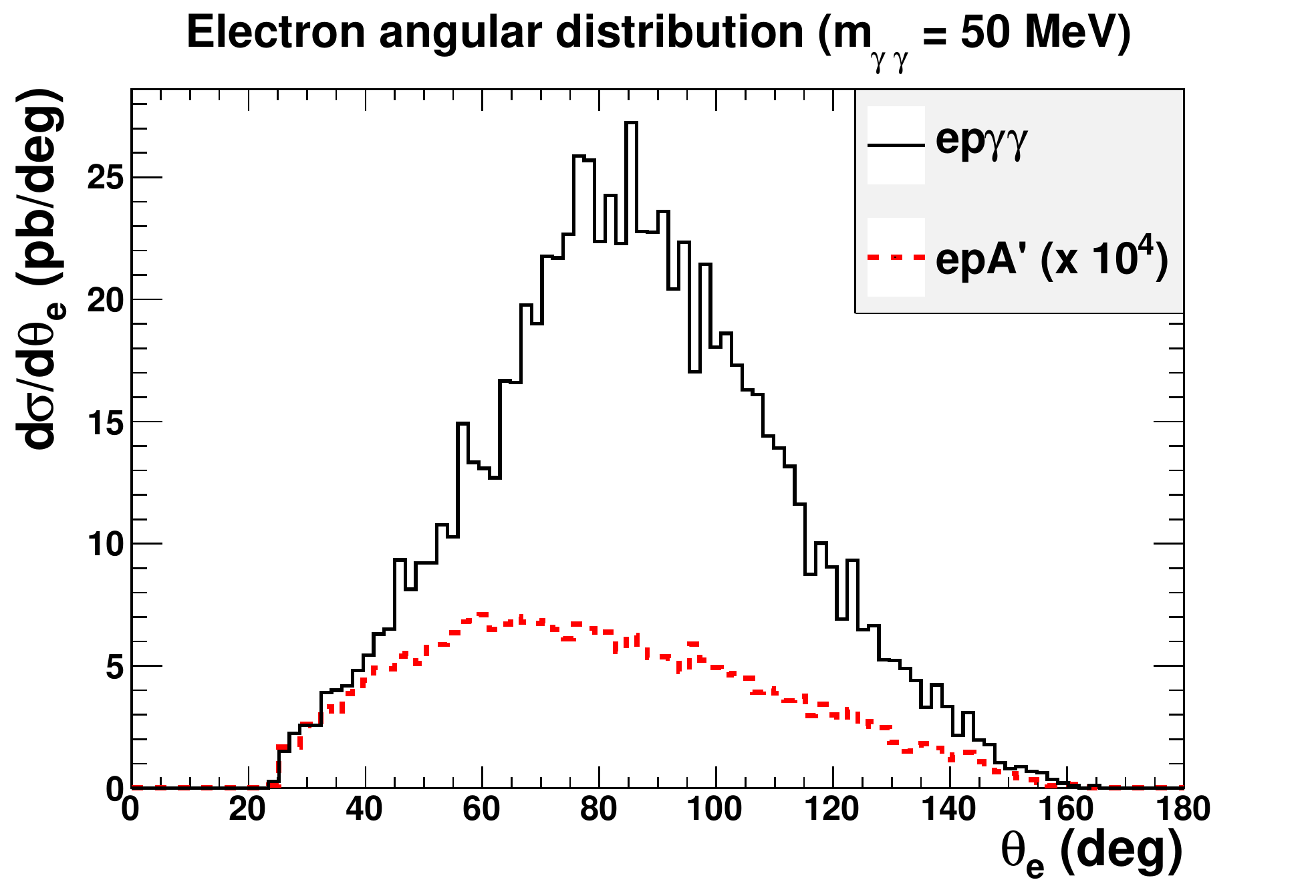} \linebreak
\includegraphics[width=0.8\columnwidth]{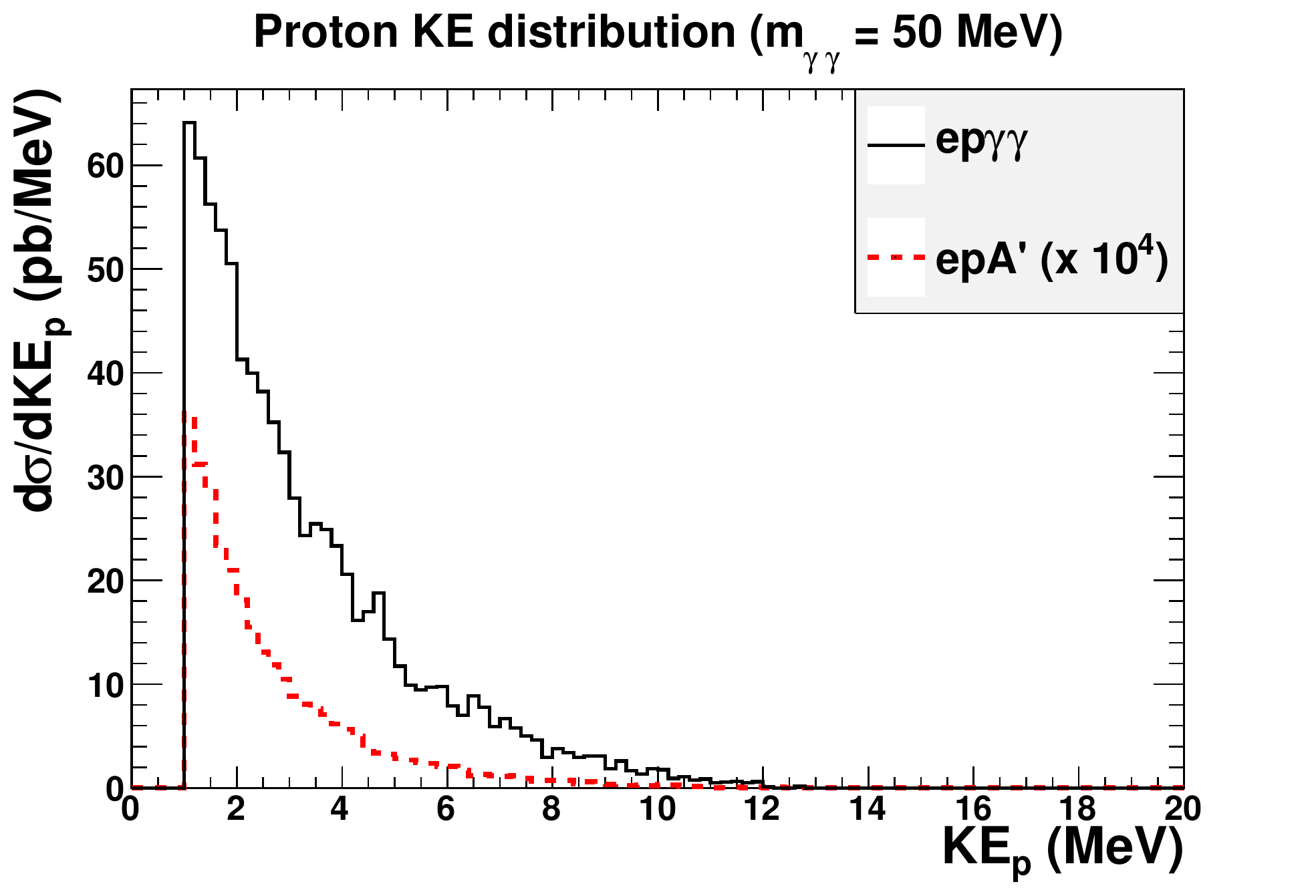} \qquad
\includegraphics[width=0.8\columnwidth]{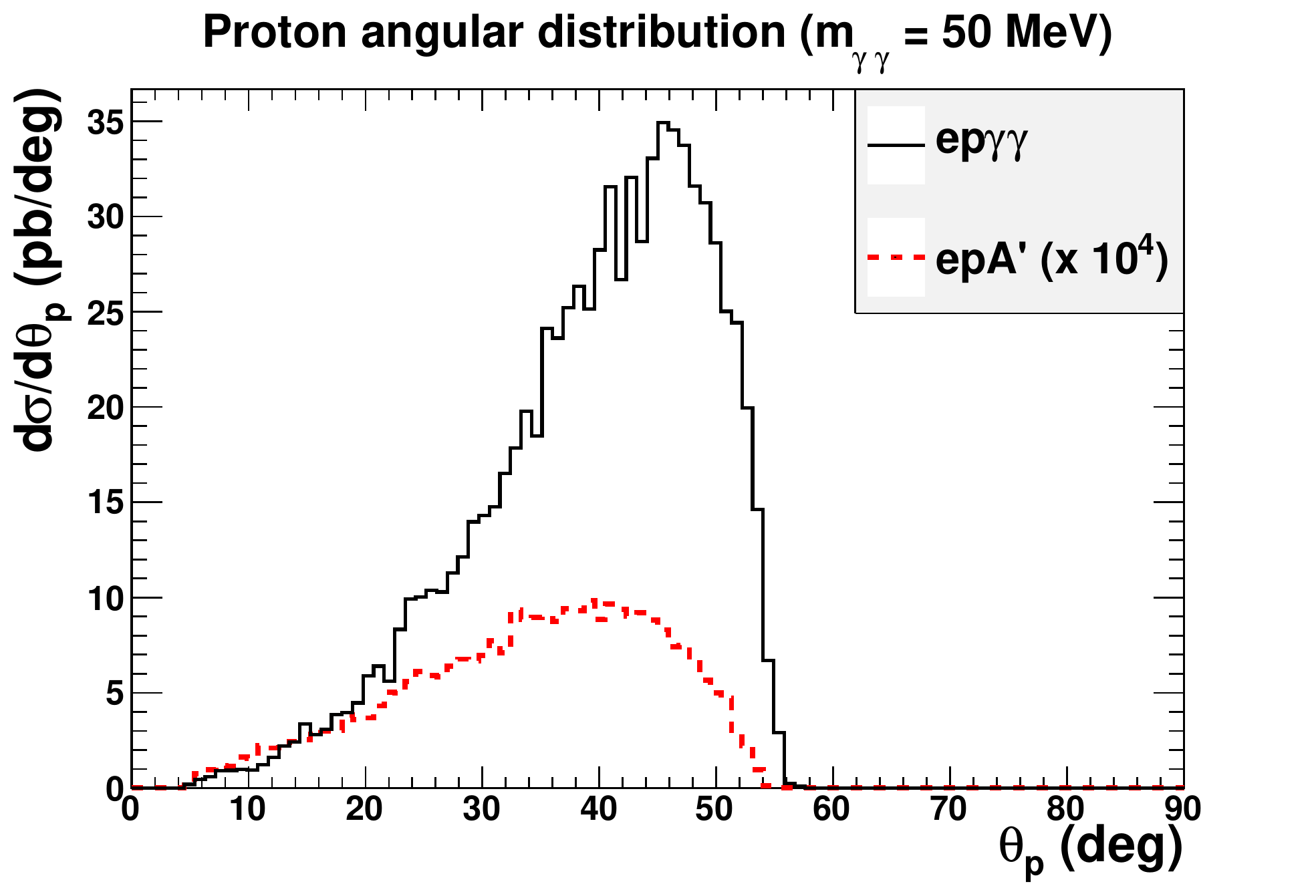} 
\caption{Final-state energy (left) and angular (right) distributions for electron (top) and proton (bottom) for the $ep\gamma \gamma$ background assuming DarkLight detector acceptances.   Here, we are restricting to an invariant mass window $m_{\gamma\gamma} = 50 \pm 1.3 \MeV$ (corresponding to $\Delta_{\cut} = 1$).   The signal distributions for $m_{A'} = 50 \MeV$ and $\alpha' = 10^{-8}$ are shown for comparison (magnified by $10^4$).}
\label{fig:GGEPDist}
\end{center}
\end{figure*}

With zero photon detection efficiency, the only irreducible background for the invisible search at ``parton-level'' (that is, not including experimental resolution effects) is two-photon bremsstrahlung $ep \to ep \gamma \gamma$; one of many contributing Feynman diagrams is shown in \Fig{fig:Feynepgg}. As shown in \Fig{fig:DiphotonMass}, the two photons have a broad invariant mass spectrum, and the differential cross section $d\sigma/dm^2_{\gamma \gamma}$ for this QED process is larger than the corresponding signal process $ep \to ep A'$ by a factor of roughly $\alphaEM^2/\alpha' \approx 10^{4}$. Note that the kinematics of the electron and proton, shown in \Fig{fig:GGEPDist}, are nearly identical to those of the signal process for a given diphoton invariant mass.  Since this is the main background, the diphoton cross section will serve as a benchmark for deciding whether additional backgrounds are negligible or not. 

As discussed in \Sec{sec:Photon}, the main goal of including photon detectors in the DarkLight setup is to make the two-photon process a reducible rather than irreducible background.  The diphoton invariant mass is given by
\begin{equation}
m^2_{\gamma \gamma} = 2 E_1 E_2 (1 - \cos \theta_{12} ),
\end{equation}
where $E_1$ and $E_2$ are the two photon energies and $\theta_{12}$ is the angle between them. This is maximized when $\theta_{12} = \pi$ and $E_1 = E_2 = E$, giving $m_{\gamma \gamma} = 2E$.  Since the DarkLight photon detectors can only see photons efficiently with $E_{\gamma} > 5 \MeV$, and since the $ep \to ep \gamma \gamma$ background is otherwise unmanageable, we set a lower limit of the signal region to $m_{A'} > 10\MeV$.

\subsubsection{One-photon and quasi-elastic}
\label{sec:Elastic}

The lowest-order QED process which generates ``missing energy'' from a radiated photon is ordinary bremsstrahlung $ep \to ep \gamma$, where quasi-elastic scattering is the limit of this process when $E_{\gamma} < 5 \MeV$. Assuming full kinematic reconstruction, the invariant mass $(p_1 + p_2 - p_3 - p_4)^2$ is zero for an $ep \gamma$ final state, since the photon has zero invariant mass.\footnote{See the comment in footnote \ref{foot:nonzero}.} So this process would not pose a background with perfect experimental resolution, even with zero photon detection.

\begin{figure}[t]
\begin{center}
\includegraphics[width=0.8\columnwidth]{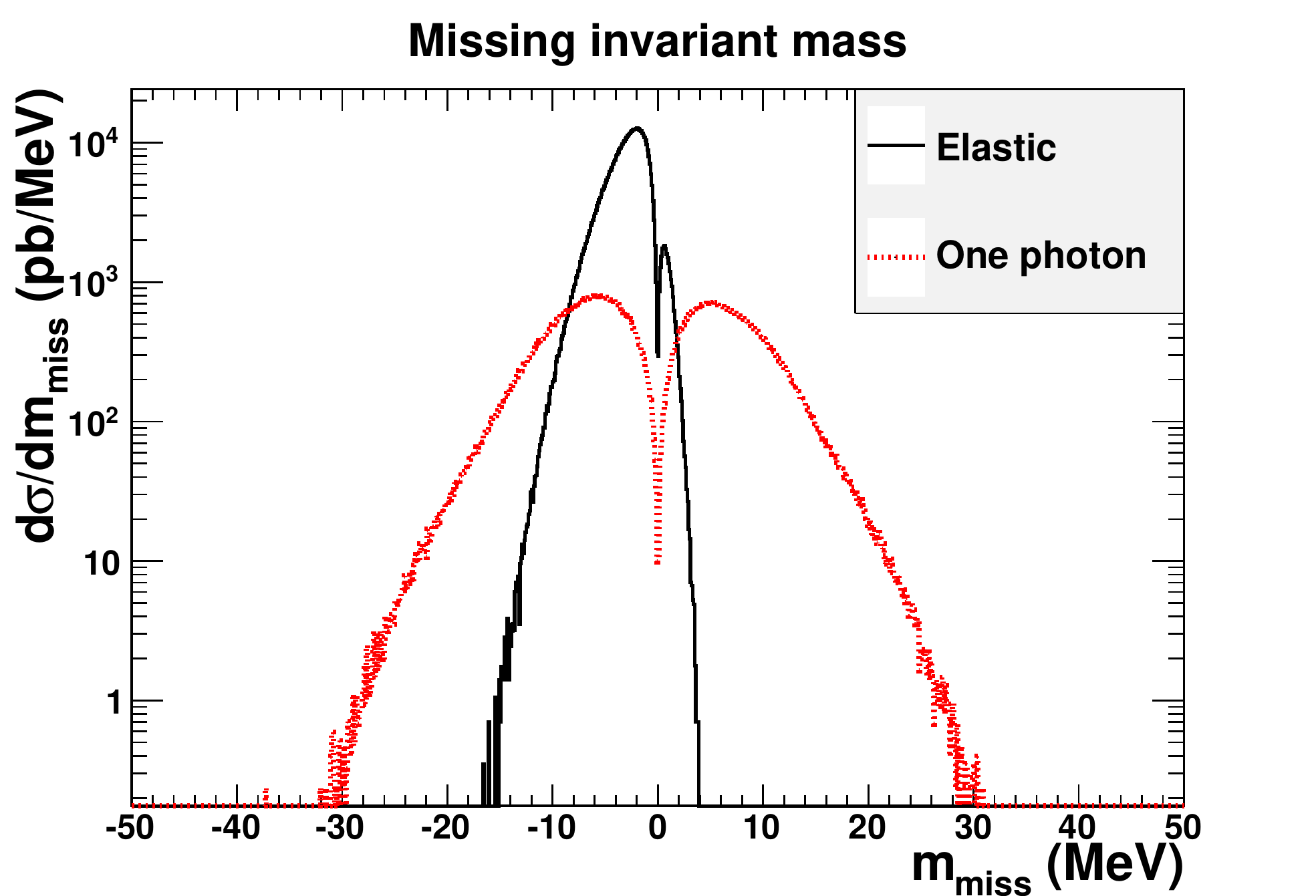}
\caption{Missing invariant mass for elastic and one-photon ($E_{\gamma} > 5 \MeV$) $ep\gamma$ events with the backwards electron cut of $\theta_{e} > 103^\circ$. The two-lobed shape is an artifact of plotting $d\sigma/dm$, rather than $d\sigma/dm^2$, where the Jacobian factor of $m$ forces the linear behavior on both sides of $m = 0$.}
\label{fig:SinglesMissingM}
\end{center}
\end{figure}

However, realistic detector resolutions on the energies, momenta, and angles of the outgoing electron and proton change this story considerably.  We can estimate the cross section of elastic and one-photon events which leak into the signal region by varying the outgoing four-vectors by independent Gaussian weights with resolutions given in \Tab{tab:DLExp}.  The result is a very broad missing invariant mass distribution shown in \Fig{fig:SinglesMissingM}, even with the backwards electron cut to be discussed further below.

It is perhaps surprising that percent-level mismeasurements of four-vector components can lead to reconstructed invariant masses on the order of 20 MeV.   The reason can be seen from \Eq{SigmaM}, which shows that $\sigma_m$ blows up as $m \to 0$.  A four-vector which should be massless is thus extremely sensitive to mismeasurement, the more so the more energy it has. Thus, the same $\Delta_{\cut}$ introduced in \Sec{sec:InvisSearch} can be used to reduce this background. We can narrow the fake invariant mass distribution further by requiring physically reasonable kinematics, namely that the missing invariant mass be sufficiently positive and the missing energy be greater than or equal to the missing invariant mass: $m_{\miss} > \DLMass$ and $1 < E_{\miss}/m_{\miss} < 1 + \Delta$. 

\begin{figure}[t]
\begin{center}
\includegraphics[width=0.8\columnwidth]{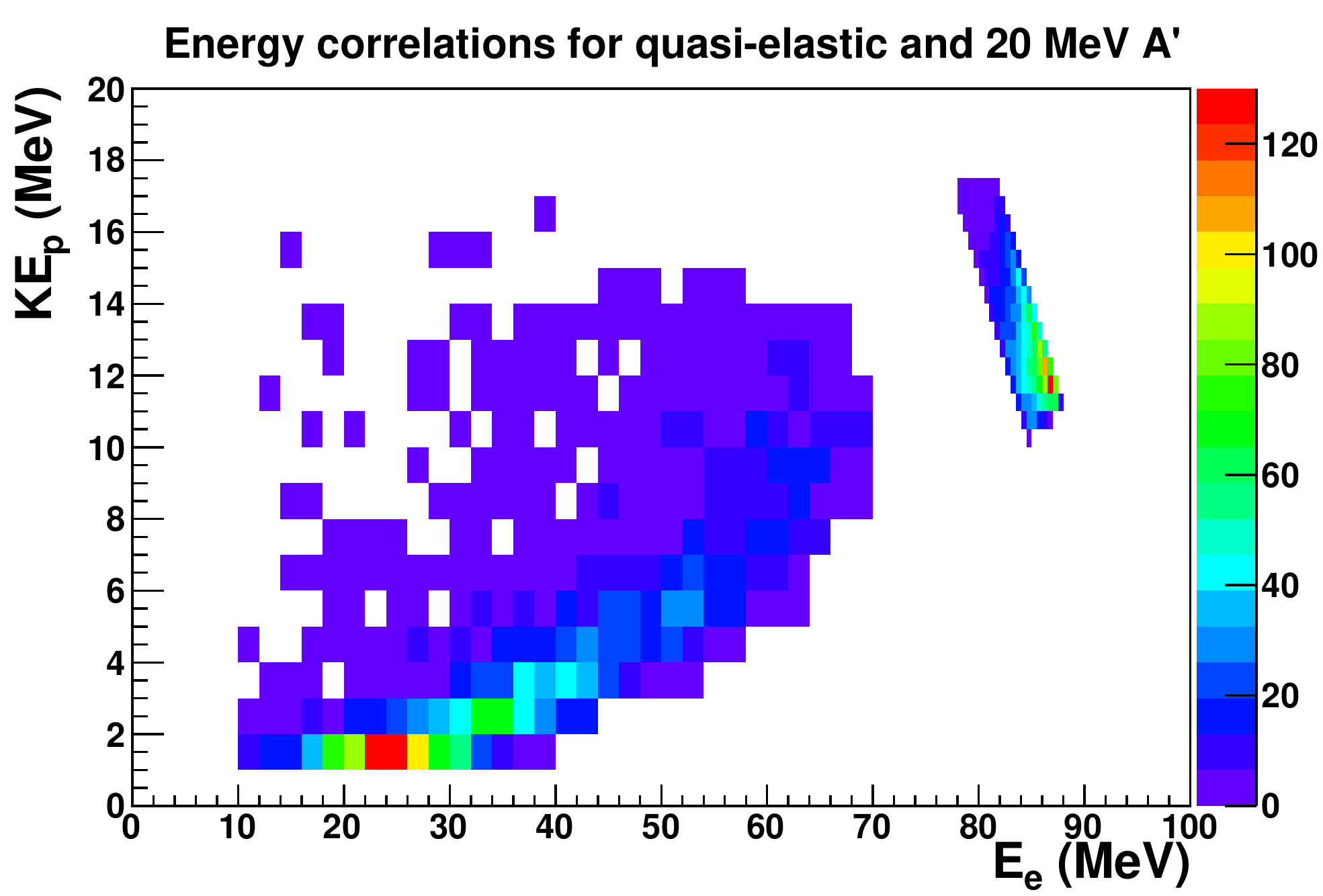}
\caption{Electron and proton energy correlations in quasi-elastic and 20 MeV $A'$ events with $\theta_{e} > 103^\circ$. The band on the right consists of quasi-elastic $ep\gamma$ events with $E_{\gamma} < 5 \MeV$, and the region on the right consists of $epA'$ events. The two regions are disjoint, showing that signal and quasi-elastic background can be distinguished kinematically.}
\label{fig:EnergyCorr}
\end{center}
\end{figure}

The background from single quasi-elastic scattering events can be suppressed almost completely by searching for the proton in the kinematic region dictated by quasi-elastic scattering.\footnote{Events in which the proton is not detected only contribute to the pileup background, since at least one proton must be present in the final state in order to form a missing invariant mass.} As shown in \Fig{fig:EnergyCorr}, even near the lower limit of the signal region $m_{A'} = 20 \MeV$, the kinematics lie in disjoint regions of phase space, so this quasi-elastic veto keeps all of the signal. For higher $m_{A'}$, the electron loses even more energy and the kinematics become even more separated, such that the quasi-elastic background cross section resulting from mis-measuring the electron or proton energies is orders of magnitude below the signal cross section for $m_{A'} > 30 \MeV$.\footnote{As we discuss in \Sec{sec:VEPP3}, this kinematic separation is also seen in the VEPP-3 experiment.}

\begin{figure}[t]
\begin{center}
\includegraphics[width=0.8\columnwidth]{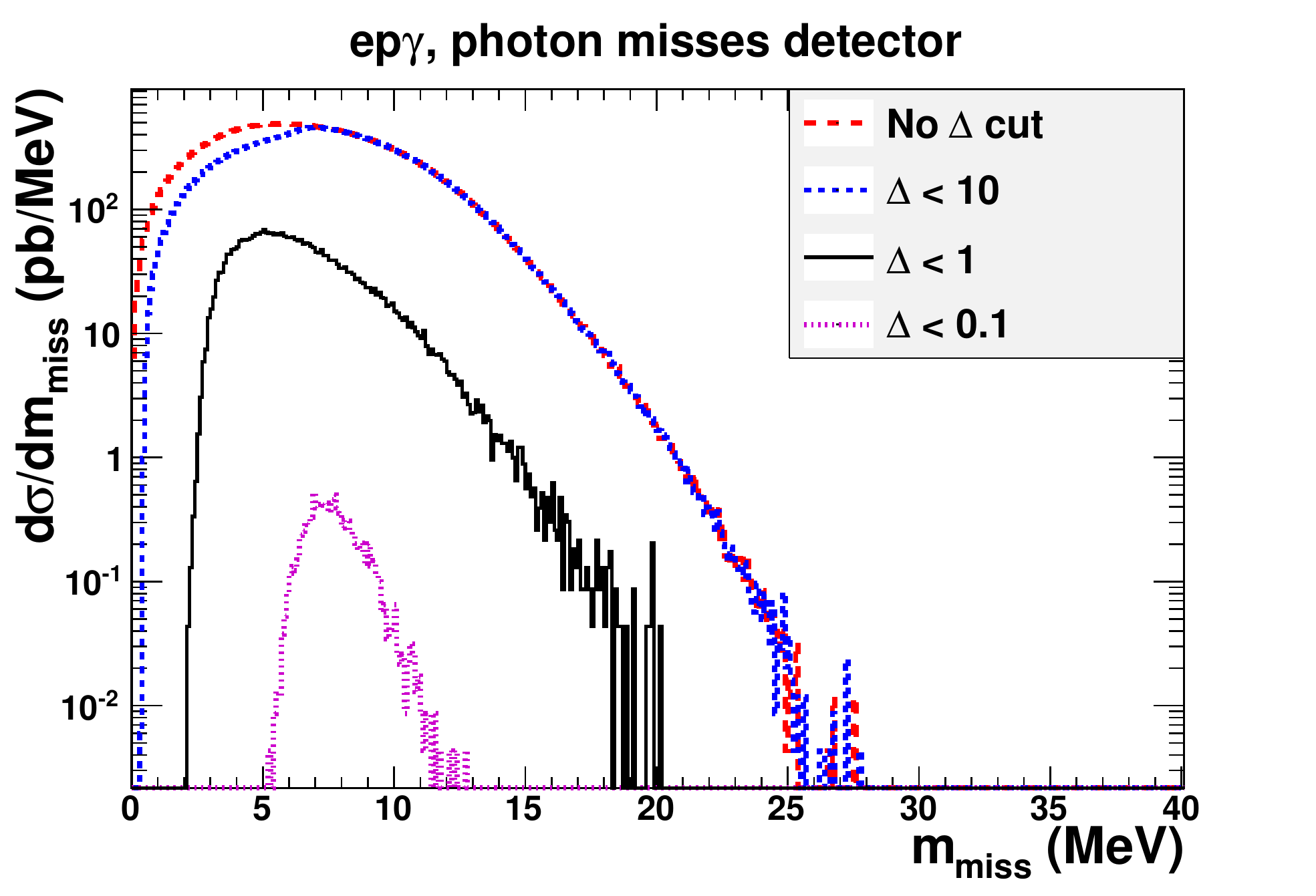} \qquad
\includegraphics[width=0.8\columnwidth]{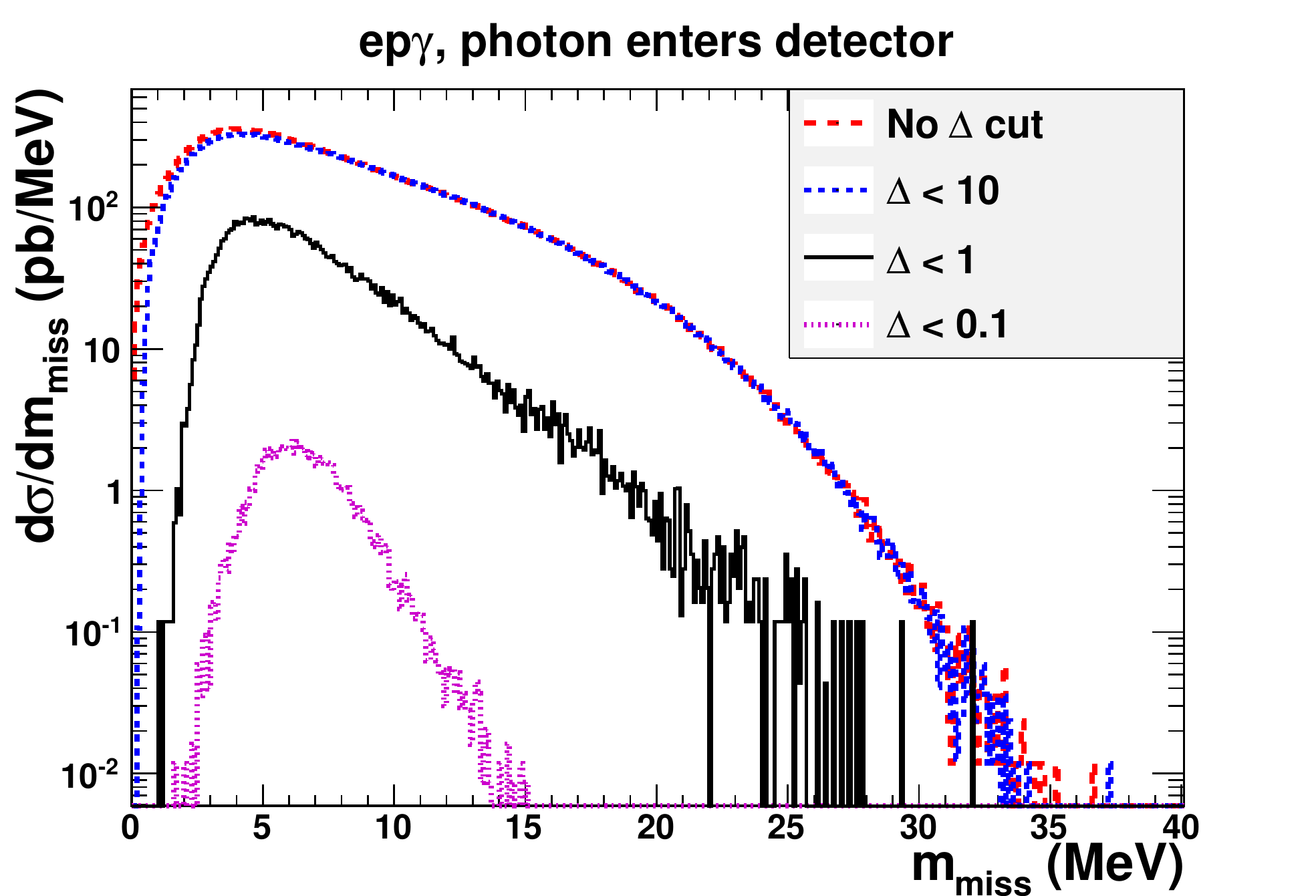} 
\caption{Measured missing invariant mass for one-photon events with $\theta_{e} > 103^\circ$, after various cuts on $\Delta$, where the photon either misses the detector (top) or enters the detector volume (bottom).}
\label{fig:OnePhotonDeltaCuts}
\end{center}
\end{figure}

On the other hand, the much wider $ep\gamma$ invariant mass distribution and the less efficient photon veto means that one-photon events cannot in general be neglected, even after a cut on $\Delta$. In \Fig{fig:OnePhotonDeltaCuts} we show the one-photon invariant mass distribution remaining after various $\Delta$ cuts.  If the photon goes undetected, an $ep \gamma$ event is indistinguishable on an event-by-event basis from  a signal event (or an $ep\gamma\gamma$ event where both photons are lost).  

\begin{table}[tdp]
\begin{center}\begin{tabular}{|c|c|} \hline \hline
\textbf{Photon eff.} & $\mathbf{\BackwardsE}$ \\
\hline
\hline 0\% & $113^\circ$ \\
\hline 50\% & $108^\circ$ \\
\hline 95\% & $103^\circ$ \\
\hline 100\% & $102^\circ$ \\
\hline \hline
 \end{tabular} \caption{Minimum allowed values of $\BackwardsE$ for a given photon detection efficiency, assuming a maximum event reconstruction rate of 50 kHz at $\DLLum$.}
\label{tab:ThetaPhoton}
\end{center}
\end{table}

In fact, the $ep \gamma$ process limits the overall event reconstruction rate for the invisible search.  DarkLight can only afford to reconstruct four-vectors at a rate of 50 kHz for the invisible search, and a backwards electron cut $\theta_e > \theta_{\cut}$ must be adjusted to keep the $ep\gamma$ rate manageable.  As can be seen from \Fig{fig:OnePhotonDeltaCuts}, the cross sections for $ep\gamma$ events where a photon is lost down the beamline is roughly the same order of magnitude as when the photon enters the detector volume.  This implies that to maintain the 50 kHz rate, the minimum allowed $\BackwardsE$ must change as a function of photon detection efficiency, as shown in \Tab{tab:ThetaPhoton}.  With 95\% photon efficiency, we gain a full $10^\circ$ in angular coverage for the same event reconstruction rate compared to no photon detection capability.  In this way, photon detection functions not only as a background-suppression technique, but also as a way to keep more of the signal cross section given a maximum event reconstruction rate.

\subsection{Pileup rejection from negative invariant mass}
\label{sec:Pileup}

An important concern in a high luminosity environment is pileup, or the presence of two or more collision events within a single timing window.  Since elastic scattering events have the same final state as signal events but an enormously higher rate, one might worry that the combinatorial background arising from mismatching an electron from one elastic event with a proton from another event in the same timing window could fake an invariant mass in the signal region $10-90\MeV$. A related issue is missing objects, for example when two elastic events occur in a timing window but one of the electrons is lost down the beam pipe, yielding a final state of one electron and two protons. Fortunately, we can efficiently deal with both of these situations simply by requiring \emph{positive} missing invariant mass-squared. 

In the notation of \Sec{sec:InvisSearch}, consider pileup between an elastic event $p_1 + p_2 \to p_3 + p_4$ and any other event $p_1 + p_2 \to p_3' + p_4' + q$ (where $q$ is invisible or not detected).  Suppose the electron from the second event is mis-reconstructed as belonging to the elastic event:
\begin{equation}
p_1 + p_2 \to p_3' + p_4.
\end{equation}
Then, since $p_1 + p_2 = p_3 + p_4$ from four-momentum conservation in the elastic event, there appears to be a missing invariant mass-squared
\begin{eqnarray}
\label{NegInvtMassFormula}
m^2 &=& (p_1 + p_2 - p_3'  - p_4)^2 \\ 
& = &(p_3 + p_4  - p_3' - p_4)^2  \nonumber\\ 
& = &(p_3 - p_3')^2 \nonumber \\ 
& = & 2m_e^2 - 2E_3 E_3' + 2\sqrt{(E_3^2 - m_e^2)(E_3'^2 - m_e^2)}\cos \theta_{3 3'}, \nonumber
\end{eqnarray}
where $\theta_{33'}$ is the angle between $p_3$ and $p_3'$. It turns out that this expression is strictly \emph{negative}, and is zero only when $p_3 = p_3'$, i.e.\ when the two events are identical. Analogous arguments hold for the other mis-reconstructed event, with $m_e$ replaced by $m_p$ and $E_3'$ replaced by $E_4'$.  So by requiring the missing invariant mass-squared to be positive, this source of background can be eliminated, up to effects arising from the finite invariant mass resolution.\footnote{This conclusion is weakened slightly because of quasi-elastic events with soft photons, but the bulk of pileup events still give a negative invariant mass-squared, and the positive tail does not extend into the signal region.}  We will make use of this trick to mitigate potential backgrounds from pileup.

To discuss the effects of pileup on the DarkLight setup, we need to compute the event rates for various processes.  The 10 ns effective timing window limits the average number of (quasi-)elastic events to $\langle N_{ep} \rangle = 0.65$, and one-photon events to $ \langle N_{ep\gamma} \rangle = 0.050$. This means we can safely assume that there will be a maximum of two $ep\gamma$ events in any timing window; as we will show in \Sec{sec:Analysis}, the effective two-event pileup cross section is already an order of magnitude below the $ep\gamma \gamma$ cross section, so higher numbers of pileup have effective cross sections which are far below $ep\gamma\gamma$ and hence negligible for this analysis. There may be several elastic events in a timing window, but by the negative invariant mass trick, these will not contribute to the signal region (up to resolution effects).

To simplify the discussion, we will assume that pileup consists of exactly two events in a timing window. This gives the following possibilities for final states:
\begin{itemize}
\item Background: $ep / ep$, $ep / ep (\gamma)$, $ep (\gamma)/ ep (\gamma)$
\item Signal: $ep / ep A'$
\end{itemize}
Here, $(\gamma)$ stands for any nonzero number of final-state photons, since as we have seen previously, one- and two-photon events are indistinguishable on an event-by-event basis due to the finite detector resolution.  As discussed in \Sec{sec:Elastic}, we can neglect $ep/ep$ because of the high proton detection efficiency, so the next leading pileup background is $ep/ep\gamma$.  To take maximal advantage of photon detection, we veto any event with a photon detected in the tracking volume, and hence do \emph{not} consider $ep(\gamma)/epA'$ as part of the signal.

\begin{figure}[tp]
\begin{center}
\includegraphics[width=0.8\columnwidth]{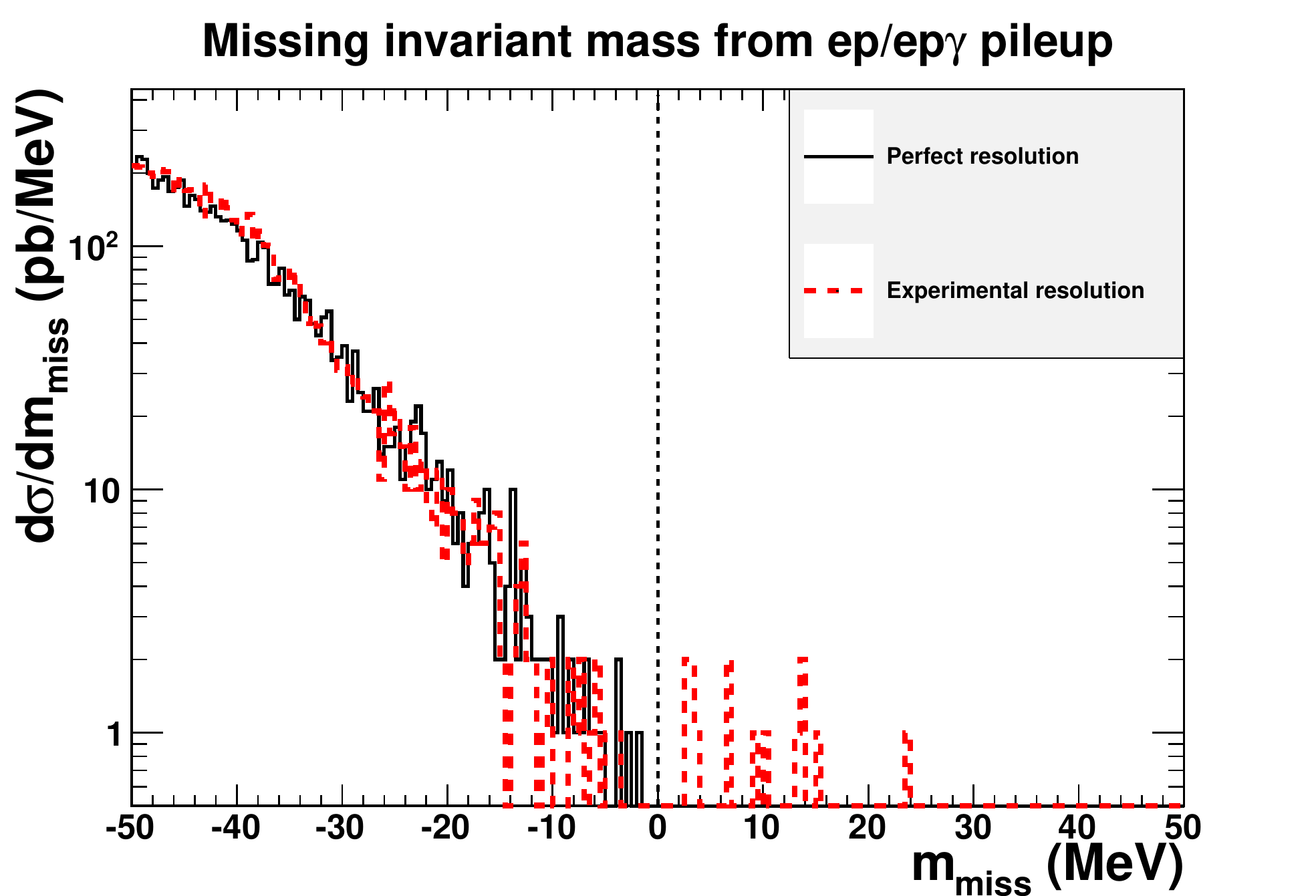}
\caption{Missing invariant mass for $ep/ep\gamma$ pileup with $\theta_{e} > 103^\circ$ for the $ep\gamma$ event. The positive tail is Monte Carlo statistics-limited, but can be effectively eliminated by an isolation veto on the trigger electron.}
\label{fig:epepgPileup}
\end{center}
\end{figure}

We now pair the backwards (trigger) electron with both protons to form two invariant masses-squared $m^2_{1}$ and $m^2_2$, and define a signal event as having one of $m_1^2$ or $m_2^2$ large and positive and the other strictly negative.\footnote{Since all the reconstructed invariant mass distributions are skewed slightly towards negative values, most of the mis-reconstructed masses for $ep / epA'$, which are ``truly'' negative by the invariant mass trick, will still be reconstructed as negative rather than slightly positive. See the discussion in \Sec{sec:Level1}. The requirement of a negative mass-squared helps eliminate background from $ep\gamma/ep\gamma$, which can have both $m^2_1 > 0$ and $m^2_2 > 0$ from either the correct or the incorrect pairing.}  The invariant mass spectra of the mis-reconstructed masses for $ep/ep\gamma$ is shown in \Fig{fig:epepgPileup}.   Assuming perfect experimental resolution, we clearly see the nonpositivity property of \Eq{NegInvtMassFormula}.  Including experimental resolution, we find a small tail which extends to positive invariant masses.  As we will show in \Sec{sec:Level1}, we can effectively eliminate this tail with an isolation veto on the trigger electron. 

This pileup mitigation technique of using negative invariant mass generalizes to higher numbers of pileup events in a straightforward way.  One can pair the trigger electron with all detected protons and require a \emph{single} positive invariant mass-squared with all the rest strictly negative.

\subsection{Other potential backgrounds}
The only truly irreducible background to the invisible search in the Standard Model is $ep \to ep Z^* \to ep \nu \overline{\nu}$ (and an analogous charged-current process), but at DarkLight energies, the $Z$ (or $W$) is so far off shell that far less than one event is expected during the entire running time of the DarkLight experiment. In the DarkLight setup, there is also a large rate of M\o ller scattering $e^- e^- \to e^- e^-$ off atomic electrons, but kinematics force the outgoing electrons to have very low energy, and these electrons are swept into the M\o ller dump shown in \Fig{fig:Detector}.

Amusingly, the DarkLight visible search background $ep \to ep e^+ e^-$ could also be a potential background for the invisible search. If all three final-state leptons are detected, we can simply impose a positron veto to discard these events. If the positron exits the detector, then the final state of $eep$ looks like a pileup event where one proton is undetected. In that case, though, the negative invariant mass trick eliminates this background entirely, since \emph{both} $e^+ e^-$ invariant masses are positive. As we will see below in \Sec{sec:Level2}, our analysis cut for dealing with pileup will require a negative invariant mass.   

Finally, the $ep e^+ e^-$ cross section for both the positron and one additional electron to escape detection is already two orders of magnitude below the $ep \gamma \gamma$ cross section where both photons miss the detector.  This exhausts the QED backgrounds to order $\alphaEM^4$. Events with three or more photons are suppressed compared to the diphoton cross section by an additional factor of $\alphaEM$, and hence can be neglected in this analysis.

\section{Analysis strategy}
\label{sec:Analysis}

Based on the considerations from the previous sections, we now present an analysis strategy for DarkLight appropriate for the invisible search.  We note that this strategy is completely different from the one presented for the visible search in \Refs{Freytsis:2009bh,PAC39}.  Essentially the only shared feature is the requirement of a backwards electron to limit the total rate to an manageable level. To take into account the cap on the total event reconstruction and data storage rates discussed in \Sec{sec:InvisSearch}, we separate our analysis cuts into object identification and vetoes (Level 1, outputting at 50 kHz) and event selection involving full event reconstruction (Level 2, outputting at 300 Hz), though we note the boundary between these regions is flexible and will ultimately depend on the structure of the readout electronics.

\subsection{Level 1: Object identification and vetoes}
\label{sec:Level1}

Because of the enormous event rates, a hierarchical trigger scheme is impractical, and so all Level 1 criteria will be applied simultaneously for the free-running trigger.  As discussed previously, the main event selection criterion is a backwards electron of angle $\theta_e > \BackwardsE$, which we label the trigger electron. This cut is designed to limit the overall event reconstruction rate, since the elastic and $ep\gamma$ rates drop sharply as a function of angle, while the signal has a fairly broad distribution of electron angles centered near $\theta_e = 90^\circ$ (see \Fig{fig:EPDist}).  Such a cut keeps approximately 10\% of the signal over most of the mass range. 

\begin{table*}[tdp]
\begin{center}
\begin{tabular}{|c||c|c|c|c|c|}\hline \textbf{Process} & \textbf{Raw (Hz)} & \textbf{Veto} & \textbf{Level 1 (Hz)} & \textbf{Level 2 (Hz)} & \textbf{50 MeV mass window (Hz)} \\
\hline 
\hline
$\boldsymbol{ep}$ & $6.5 \times 10^{7}$ & $p$ & $< 1$ & $< 10^{-2}$ & $< 10^{-3}$  \\
\hline $\boldsymbol{ep\gamma}$ & $5.0 \times 10^{6}$ & $\gamma$ & $5.0 \times 10^{4}$ & $2. 0 \times 10^2$  & $< 10^{-3}$ \\
\hline $\boldsymbol{ep\gamma \gamma}$ & $1.6 \times 10^5$ & $\gamma$ & $1.7 \times 10^2$ &  77 & 2.4\\
\hline $\boldsymbol{epe^+e^-}$ & $6.6 \times 10^3$ & $e^-, e^+$ & $1.2 \times 10^2$ & 2.3 & $< 10^{-2}$ \\
\hline $\boldsymbol{ep/ep}$ & $3.1 \times 10^7$ & $p,e^-$ &$1.2 \times 10^3$ &  $< 10^{-2}$ & $< 10^{-3}$ \\
\hline $\boldsymbol{ep/ep\gamma}$ & $2.4 \times 10^6$ & $p,\gamma,e^-$ &$4.1 \times 10^2$&  $< 10^{-2}$ &  $< 10^{-3}$\\
\hline $\boldsymbol{ep\gamma/ep\gamma}$ &  $2.4 \times 10^5$ & $\gamma,e^-$ & $1.3 \times 10^3$&  7.3 & 0.27\\
\hline
\hline \textbf{Total Background} & $7.1 \times 10^7$ & --  & $5.0 \times 10^4$ & $2.8 \times 10^{2}$ & 2.7 \\
\hline
\hline \textbf{Signal} &  $5.4 \times 10^{-2}$ &  none & $1.3 \times 10^{-2}$ & $1.3 \times 10^{-2}$ & $9.8 \times 10^{-3}$ \\
\hline
 \end{tabular} 
\caption{Example rates for a 50 MeV $A'$ search, assuming a luminosity of $\DLLum$, $\alpha' = 3 \times 10^{-8}$, 95\% photon detection efficiency, $\BackwardsE = 103^\circ$, and $\Delta_{\cut} = 1$ (1.3 MeV invariant mass resolution). The singles rates are total rates and include the pileup rates.  The $ep/ep$ and $ep/ep\gamma$ rates include an arbitrary number of elastic events. The vetoes $p$, $\gamma$, $e^-$, and $e^+$ refer to the four types of Level 1 vetoes described in \Sec{sec:Level1}. Although $\Delta_{\cut} = 1$ does not affect the 50 MeV signal, mis-measurement forces a small portion of the cross-section to lie outside the 50 MeV mass window, hence the reduction in signal rate after the mass window cut.}
\label{tab:ExampleRates}
\end{center}
\end{table*}

In addition to the backwards electron, we impose a veto on any of the following situations:
\begin{itemize}
\item Elastic $p$ veto:  A proton consistent with quasi-elastic kinematics (i.e.~coplanar with the trigger electron, and with energy and angle correlated to that of elastic scattering within experimental resolutions);
\item Any $\gamma$ veto:  One or more detected photons above $5~\MeV$;
\item Nearby $e^-$ veto:  Any additional electrons with less than $5^\circ$ separation in $\theta$ from the trigger electron;
\item Any $e^+$ veto: Any positron with $p_T > 10 \MeV$.
\end{itemize}
These vetoes are all 1-bit pieces of information (the first veto can be applied using a look-up table based on quasi-elastic kinematics), and so can be applied at readout time. Note that the backwards electron cut is correlated with photon efficiency as in \Tab{tab:ThetaPhoton}. 

The first veto suppresses the elastic background except in cases where the detector fails to register the proton.  The second veto suppresses the background from $ep(\gamma)$ events.\footnote{We assume for this analysis that the rate of electrons faking photons is zero; relaxing this assumption only increases the probability of a veto, and does not contribute to any additional background.} The third veto eliminates the small positive invariant mass tail of elastic pileup, which arises from events whose ``true'' fake invariant mass is very close to zero.  From \Eq{NegInvtMassFormula}, a near-zero invariant mass from $ep/ep$ pileup only occurs when the paired events have very similar kinematics.  Requiring all additional electrons to be at least $5^\circ$ separated from the trigger electron forces any accompanying protons from underlying elastic events to have sufficiently different kinematics that the reconstructed invariant mass remains negative.\footnote{In fact, we can in principle have any number of additional elastic events accompanying a signal event.  After applying the third veto, essentially all the mis-paired invariant masses will be negative even with finite experimental resolution, allowing us to take full advantage of the negative invariant mass trick.  Note that we could improve the third veto by requiring only separation in opening angle $\Omega$ between the two electrons instead of $\theta$, but the elastic rates at these large $\theta$ angles are sufficiently low that the improvement would be modest at best.}  Finally, the fourth veto ensures that the $epe^+e^-$ background does not contribute to the invisible search if the positron is detected.

The rates after these selections are shown in \Tab{tab:ExampleRates}. Pileup rates are calculated by assuming a trigger on a backwards electron, then calculating the probability for seeing a second event of the given type within the following 10 ns window according to Poisson statistics and weighting the cross section by this probability. We see that the output of Level 1 is 50 kHz as desired. 

\subsection{Level 2: Event reconstruction and selection}
\label{sec:Level2}

\begin{figure}[tp]
\begin{center}
\includegraphics[width=0.8\columnwidth]{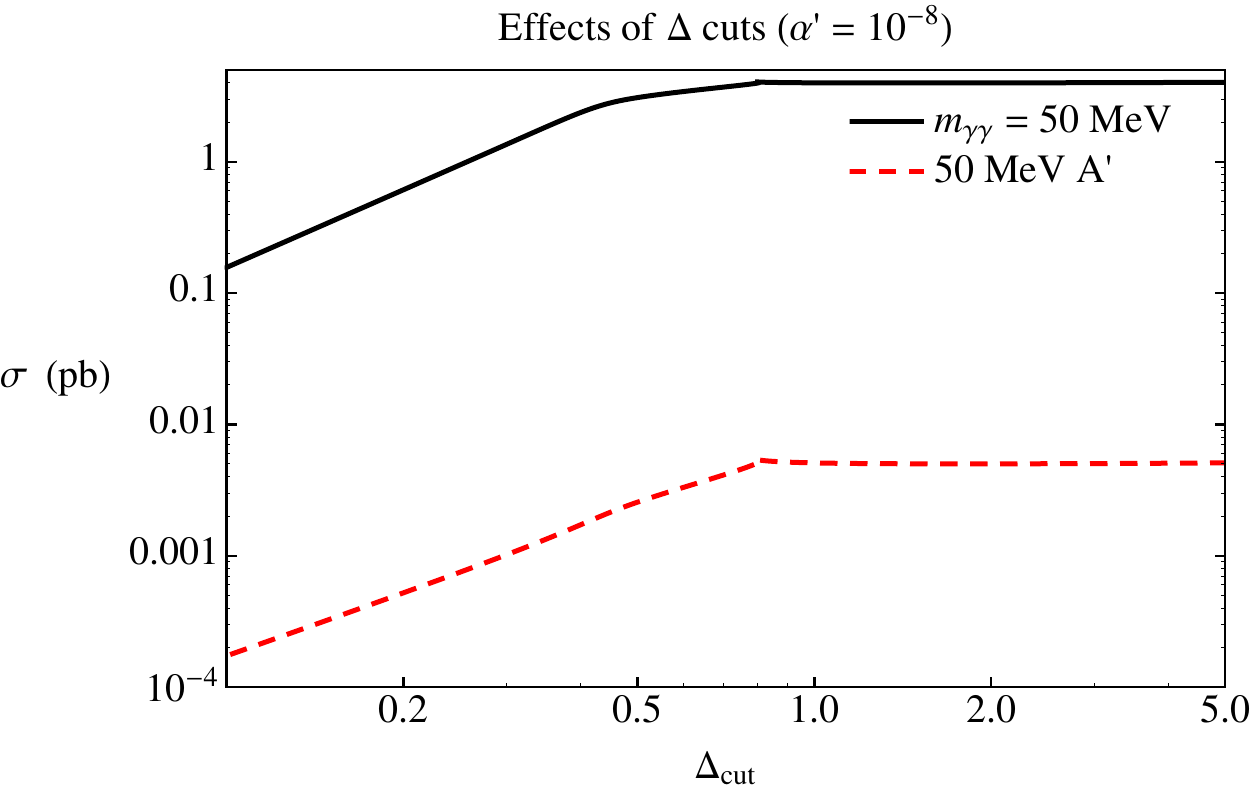}
\caption{Signal and background cross sections for $m_{A'} = 50 \MeV$ as a function of $\Delta_{\cut}$.}
\label{fig:DeltaCuts}
\end{center}
\end{figure}

After the Level 1 vetoes, we are able to perform full event reconstruction at 50 kHz, enabling additional Level 2 cuts for signal discrimination to yield a final data storage rate of 300 Hz.  From \Tab{tab:ExampleRates}, the dominant backgrounds after Level 1 are single $ep(\gamma)$ events and $ep\gamma/ep\gamma$ pileup.  The $ep\gamma$ events typically have small values of $m_{\miss}^2$ and do not enter the signal region, leaving mostly $ep\gamma \gamma$. The $ep\gamma/ep\gamma$ pileup can be largely eliminated using the negative invariant mass trick of \Sec{sec:Pileup}.  We can also improve the signal invariant mass resolution using the $\Delta_{\cut}$ criteria of \Sec{sec:InvtMassRes}.

At Level 2, we pairwise combine the trigger electron with the various detected protons, and calculate all invariant masses.  We keep only events with:
\begin{itemize}
\item A single invariant mass-squared greater than $(\DLMass)^2$;
\item All other invariant mass-squared pairs negative; 
\item $\Delta < \Delta_{\cut}$.
\end{itemize}
The $\Delta < \Delta_{\cut}$ selection further reduces the tail on the $ep(\gamma)$ singles and pileup, and should be optimized to maximize the signal to background ratio for a desired invariant mass resolution.\footnote{For this reason, it may be advantageous to apply the $\Delta$ cut offline.}  In \Fig{fig:DeltaCuts}, we illustrate the effect of $\Delta$ cuts on the signal and $ep\gamma \gamma$ background cross sections for $m_{A'} = 50 \MeV$ and 95\% photon efficiency; both cross sections drop at the same rate as $\Delta_{\cut}$ is decreased, so the reach is in fact worsened by stronger cuts on $\Delta$. On the other hand, as shown previously in \Tab{tab:MassRes}, the mass resolutions increase uniformly with stronger cuts on $\Delta$. We note again that $\Delta_{\cut} = 10$ is the maximum kinematically allowed value of $\Delta$ for the lightest $A'$ in the signal region, $m_{A'} = 10 \MeV$, so applying this cut keeps 100\% of the signal.  We will consider some representative values of $\Delta_{\cut} = 10, 1, 0.5$ in the analysis below.  

As seen in \Tab{tab:ExampleRates}, the only substantial backgrounds remaining after the Level 2 criteria are $ep\gamma$ and $ep\gamma\gamma$, with final data storage rate below the 300 Hz target.  For $m_{A'} > 20 \MeV$, $ep\gamma \gamma$ dominates.  These background rates could be further reduced by improving the photon detection efficiency.

\subsection{Analysis example}
\label{sec:ExampleRates}

\begin{figure}[tp]
\begin{center}
\includegraphics[width=0.8\columnwidth]{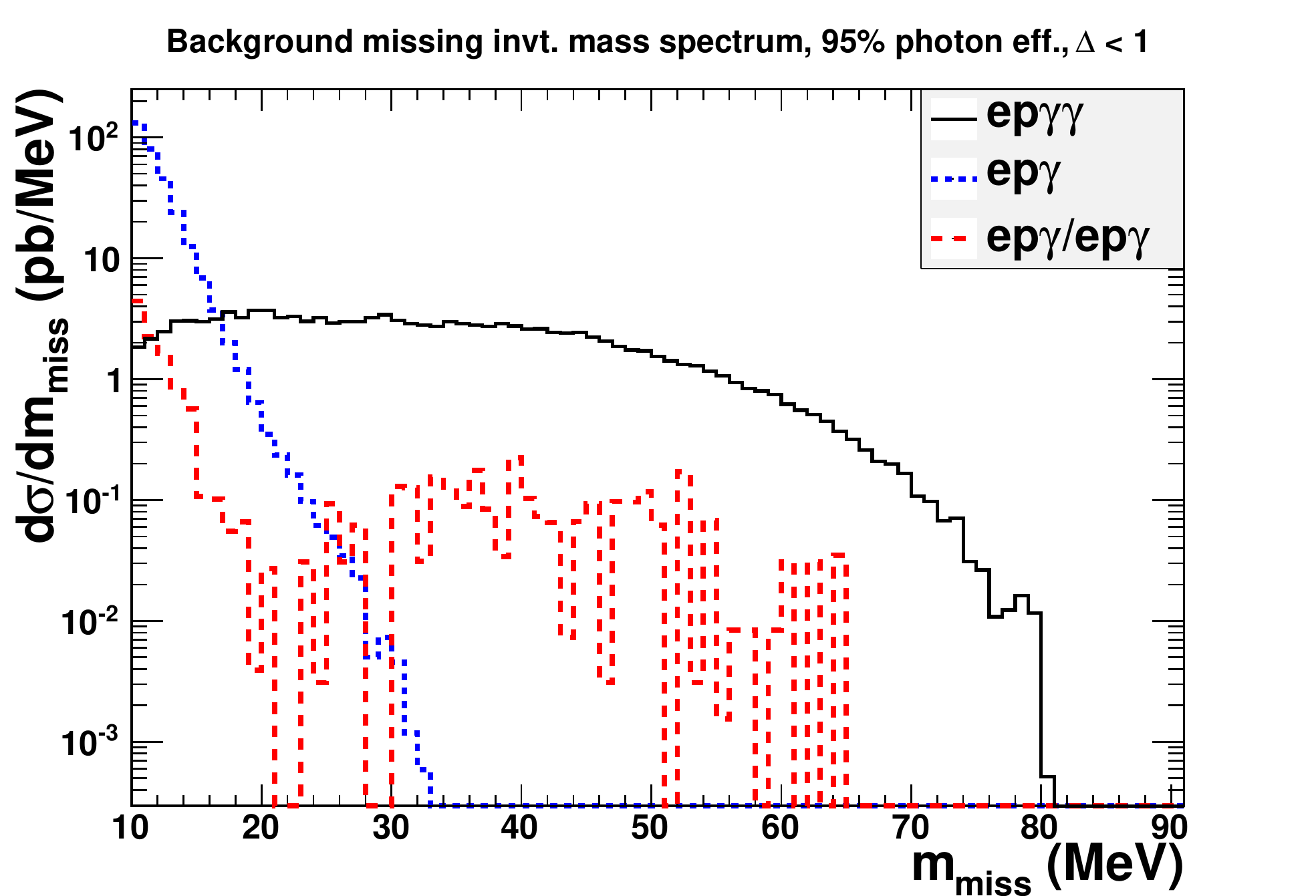}
\caption{Background missing invariant mass distributions after Level 2 criteria.  The structures in $ep\gamma/ep\gamma$ are due to limited Monte Carlo statistics.}
\label{fig:BGSpectrum}
\end{center}
\end{figure}

Here, we demonstrate the above analysis strategy by applying these cuts to an example of a search for a 50 MeV $A'$ boson with $\alpha' = 3 \times 10^{-8}$. We take 95\% photon detection efficiency, the corresponding $\BackwardsE = 103^\circ$, and $\Delta_{\cut} = 1$, giving an invariant mass resolution of 1.3 MeV.  In \Fig{fig:BGSpectrum}, we show the missing invariant mass spectrum of the various background processes after the analysis cuts, confirming that the $ep\gamma \gamma$ cross section dominates for moderately large $m_{\miss}$.  For pileup backgrounds, the elastic proton veto suppresses $ep/ep$ and $ep/ep\gamma$ somewhat at Level 1, but the negative invariant mass trick suppresses the Level 2 rates to well below the diphoton cross section, so the only relevant pileup background is $ep\gamma/ep\gamma$ (which is still an order of magnitude smaller than $ep\gamma \gamma$ in the signal region).  The last column of \Tab{tab:ExampleRates} shows the rate for the relevant background processes in the $50 \pm 1.3\MeV$ invariant mass bin.   Even with the relatively strict selection criteria, the signal rate is over two orders of magnitude smaller than the background, highlighting the need for a high luminosity experiment.

\begin{figure}[tp]
\begin{center}
\includegraphics[width=0.8\columnwidth]{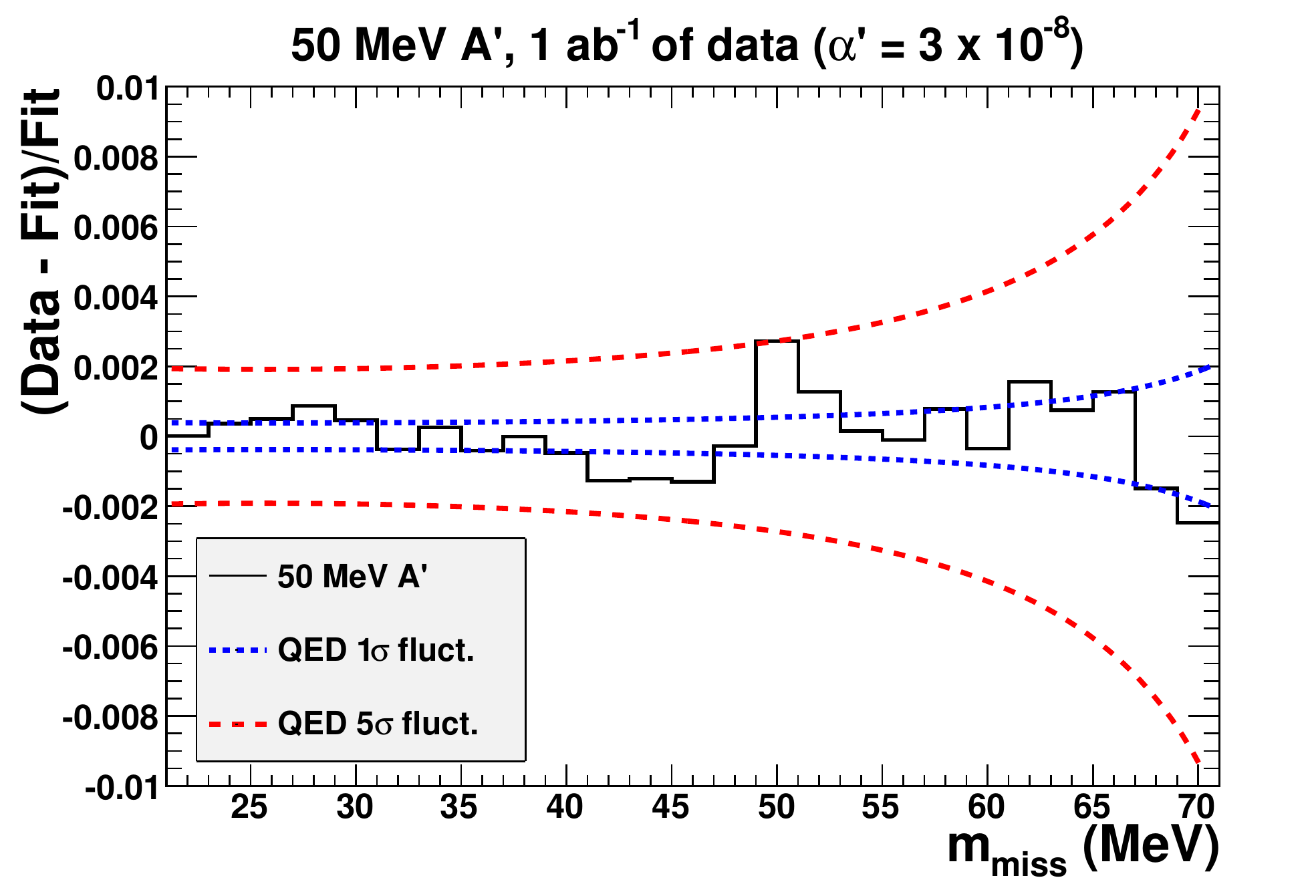}
\caption{Results of a sample analysis with 1 ab$^{-1}$ of data. The resonance at $m_{A'} = 50 \MeV$ is clearly visible after subtracting a smooth background fit shape.}
\label{fig:BumpPlot}
\end{center}
\end{figure}

To illustrate what a discovery of a 50 MeV $A'$ with $\alpha' = 3 \times 10^{-8}$ would look like, we generated 1 ab$^{-1}$ of pseudodata (corresponding to 60 days of live data taking) including the signal and all the backgrounds mentioned above.   We fit to the full missing invariant mass spectrum for $m_{\miss} > 20 \MeV$ using a smooth sixth-order polynomial.  For the case of background only, this polynomial should match the measured distribution up to statistical fluctuations.  In the presence of a signal, a plot of data minus fit divided by fit will show a spike at a candidate $m_{A'}$ value.  This is demonstrated in \Fig{fig:BumpPlot}, which shows a clear spike at $m_{\miss} = 50 \MeV$ which lies above the expected $5\sigma$ band from background statistical fluctuations.  Hence, with an inverse attobarn of data, one can resolve the faint signal of an invisible $A'$ vector boson despite the large backgrounds.

\section{Experimental reach}
\label{sec:Reach}

\subsection{DarkLight reach}
\label{sec:DLReach}

\begin{figure*}[tp]
\begin{center}
\includegraphics[width=0.8\columnwidth]{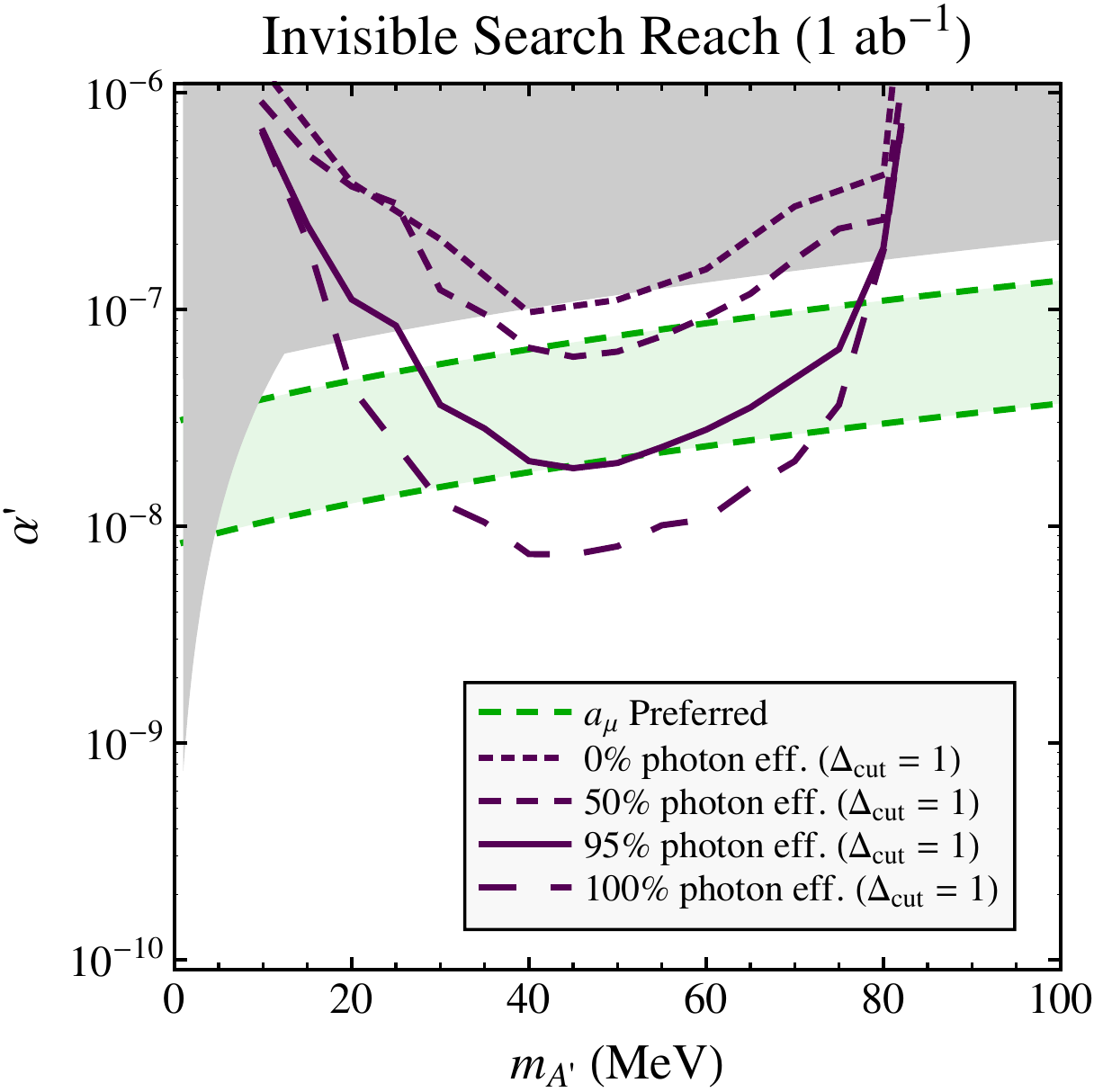} \qquad
\includegraphics[width=0.8\columnwidth]{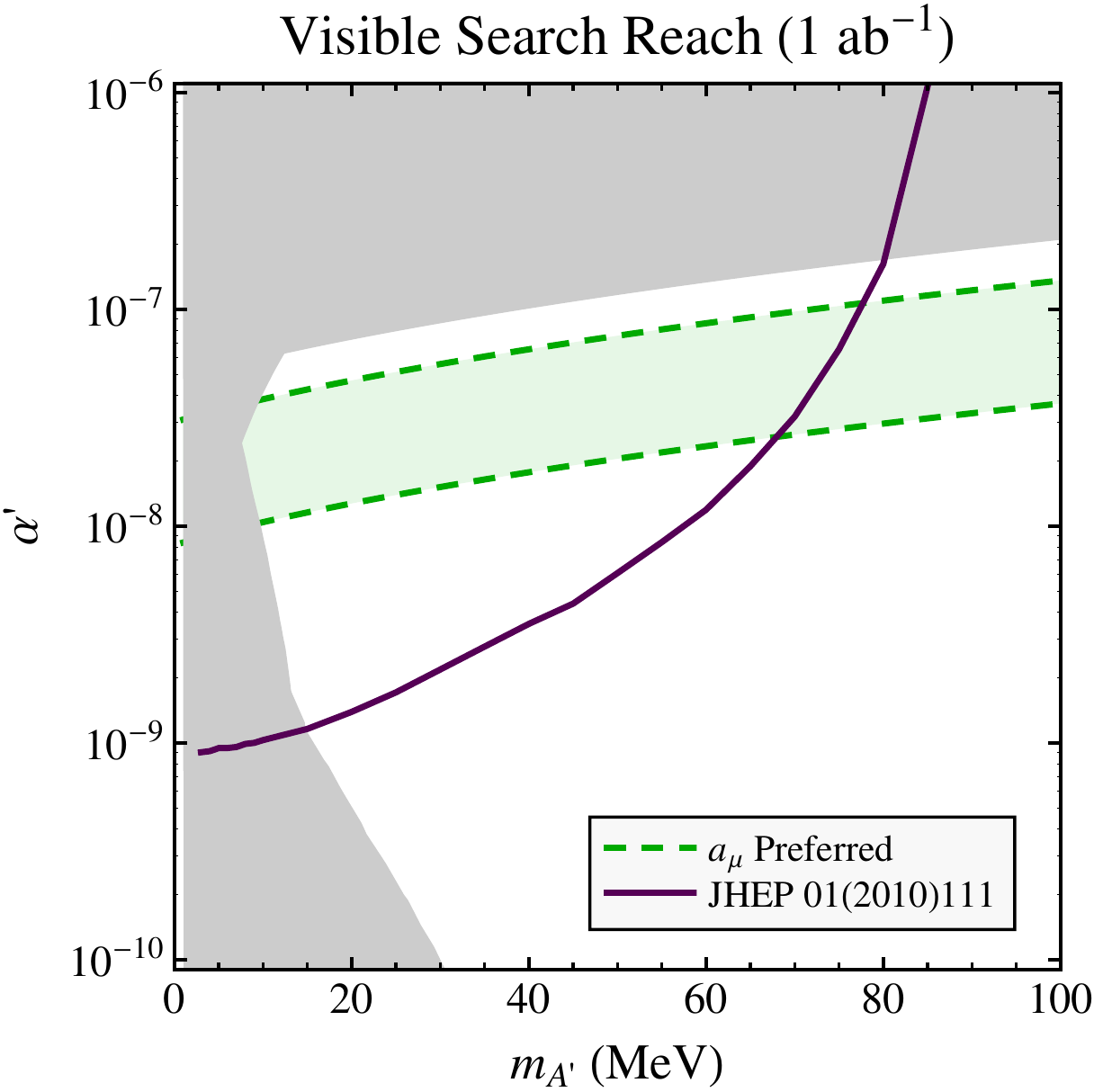}
\caption{Left: DarkLight invisible search reach for various photon efficiencies. The gray shaded area indicates constraints from anomalous magnetic moment measurements, with the green region indicating the ``welcome'' region where an $A'$ could explain the $(g-2)_\mu$ discrepancy.  Right: The visible search reach is shown for comparison (adapted from \Ref{Freytsis:2009bh}) and includes additional constraints from beam dump experiments. The fluctuations in the reach for high photon efficiency are an artifact of the difficulty of achieving high enough Monte Carlo statistics for events where both photons miss the detector.}
\label{fig:DLReachD1}
\end{center}
\end{figure*}

\begin{figure}[tp]
\begin{center}
\includegraphics[width=0.8\columnwidth]{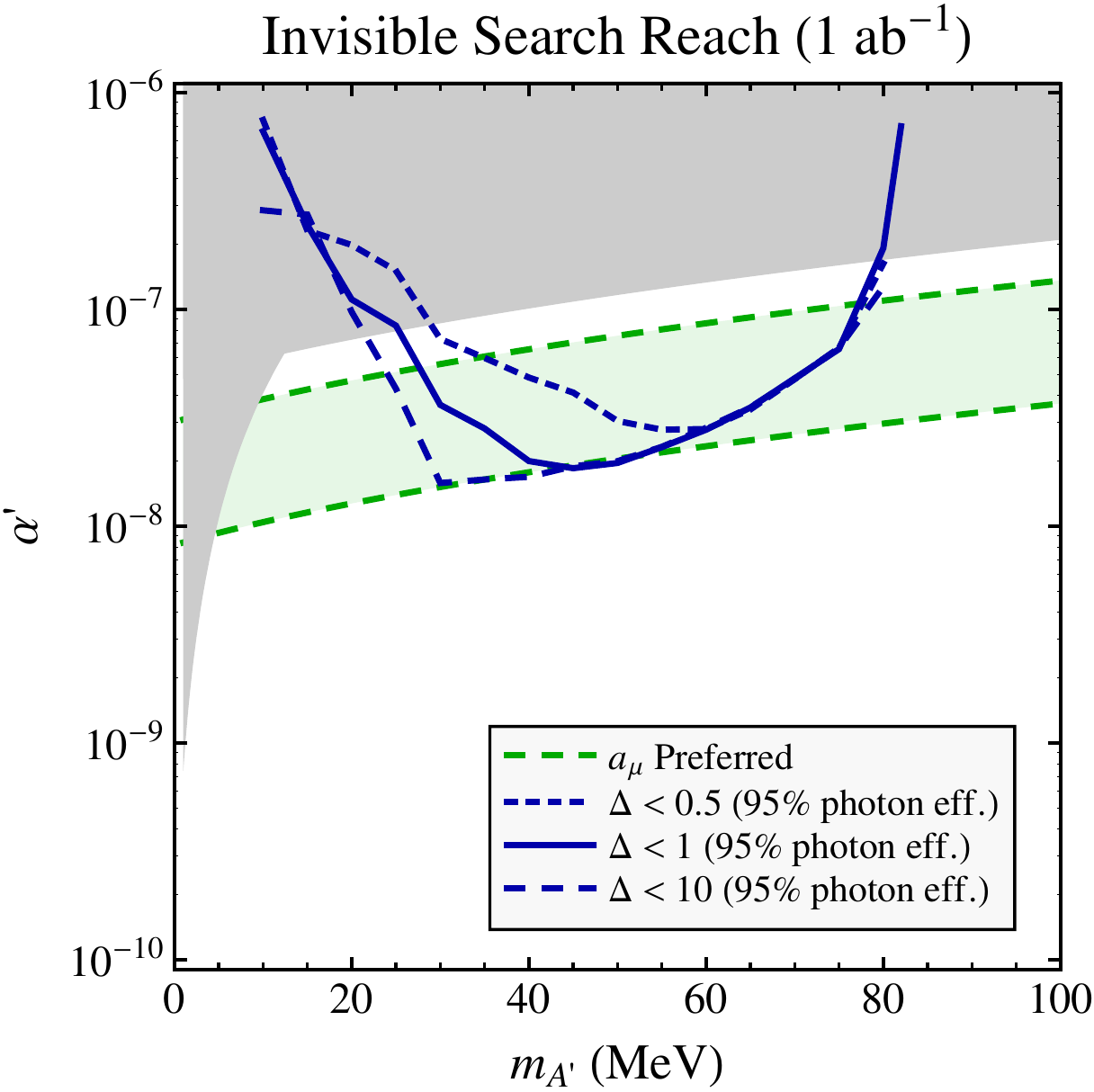}
\caption{DarkLight invisible search reach with 95\% photon efficiency for various choices of $\Delta_{\cut}$.}
\label{fig:DLReachP95}
\end{center}
\end{figure}

Following the procedure outlined in \Sec{sec:ExampleRates}, and requiring a $5\sigma$ signal-to-background significance (that is, $S/\sqrt{B} > 5$, where $S$ and $B$ are the number of signal and background events respectively), we find the potential DarkLight invisible reach shown in \Fig{fig:DLReachD1}. We chose four representative values of photon efficiency: $0\%$, $50\%$, $95\%$, and $100\%$, with the appropriate value of $\BackwardsE$ from \Tab{tab:ThetaPhoton}.  For comparison, we also show the DarkLight visible ($A' \to e^+ e^-$) reach from the study in \Ref{Freytsis:2009bh}.  The shaded areas represent excluded regions of parameter space from constraints on the electron and muon anomalous magnetic moments \cite{Fayet:2007ua, Pospelov:2008zw}, or from beam dump experiments \cite{Riordan:1987aw, Bross:1989mp} in the case of the visible search. Note that the beam-dump experiments, which place strong constraints on long-lived $A'$ particles and exclude a region of parameter space with small $\alpha'$ for the visible search, do not place any constraints on the invisible search. The green region represents the discrepancy between experimental and theoretical values for $(g-2)_\mu$, such that the discovery of an $A'$ with mass and coupling within the band could explain the discrepancy \cite{Pospelov:2008zw}.

We see that with 95\% photon detection efficiency, we can probe the majority of the preferred region.   With no photon detection, the invisible search does not even extend to a region of parameter space which is not already excluded by anomalous magnetic moment data.  Even with 50\% photon efficiency, we can barely probe the preferred region, thus showing the importance of efficient photon detection.  In \Fig{fig:DLReachP95}, we show the effect of $\Delta$ cuts on the reach for $95\%$ photon efficiency.  As shown previously in \Fig{fig:DeltaCuts}, although the invariant mass resolution improves with tighter cuts on $\Delta$, the loss of signal cross section outweighs this improvement and causes the reach to decrease. However, taking $\Delta_{\cut} = 1$ gives mass resolutions on the order of $1-2 \MeV$ throughout the entire accessible mass range with only a modest loss of sensitivity in the low-mass region.

\subsection{Potential improvements}
\label{sec:Improve}

There are two straightforward ways to potentially improve the DarkLight invisible search by modifying the detector apparatus:  increase photon efficiency or increase photon acceptance.  The importance of photon efficiency is already highlighted in \Fig{fig:DLReachD1}, where there is a potential factor of 3 gain in going from 95\% to 100\% efficiency, and part of this gain could be achieved by using a thicker photon detector.   On the other hand, in our studies we find that once $25^\circ$ -- $165^\circ$ photon angular coverage is achieved, only modest gains are possible by increasing photon acceptance in the forward direction.  In principle, one could add a forward photon endcap to enable forward coverage down to $15^\circ$, a limit which is set by the forward lead collimator shown in \Fig{fig:Detector}.\footnote{As mentioned in \Sec{sec:Photon} the backwards angle $165^\circ$ is limited by the beam collimator.}  However, after all of the analysis cuts have been applied, the number of background events with photons in the $15^\circ$ -- $25^\circ$ range is proportionally small, since most of the remaining backgrounds involve one photon traveling within $5^\circ$ of the beamline, a region that can never be instrumented.  

A more challenging way to improve the DarkLight invisible search would be to increase signal acceptance.  Our Level 1 trigger strategy did not require any calculation of invariant masses, since all cuts involving invariant masses were only applied at Level 2 after event reconstruction.  This meant that the 50 kHz output rate at Level 1 was dominated by $ep\gamma$ events which were almost entirely thrown out after the Level 2 signal discrimination cuts.  If we were able to focus on a particular invariant mass region at Level 1, we could significantly loosen the backwards electron cut, thus keeping more of the signal.  However, that would require extremely fast event reconstruction at better than 50 kHz to improve the reach.

\subsection{Comparison to VEPP-3 proposal}
\label{sec:VEPP3}

It is instructive to compare the DarkLight invisible search strategy advocated here to the recent VEPP-3 proposal in \Ref{Wojtsekhowski:2012zq}.  The VEPP-3 positron storage ring can deliver a high-current positron beam of energy 500 MeV.  When the beam is incident on a hydrogen target, the process $e^+ e^- \to \gamma A'$ can allow a search for $A'$ bosons.  This search is independent of the $A'$ decay mode, since one can compute the $A'$ invariant mass using just the single final-state photon, which is detected using two symmetrical calorimeters covering $1.5^\circ < \theta < 4.5^\circ$.  The center-of-mass energy for a 500 MeV beam is about 22.6 MeV, and so the VEPP-3 experiment is designed to search for an $A'$ with mass $5-20\MeV$. This is complementary to DarkLight, which as shown in \Fig{fig:DLReachD1} is sensitive to $A'$ masses of $20-80 \MeV$.  

\begin{figure}[t]
\begin{center}
\includegraphics[width=0.8\columnwidth]{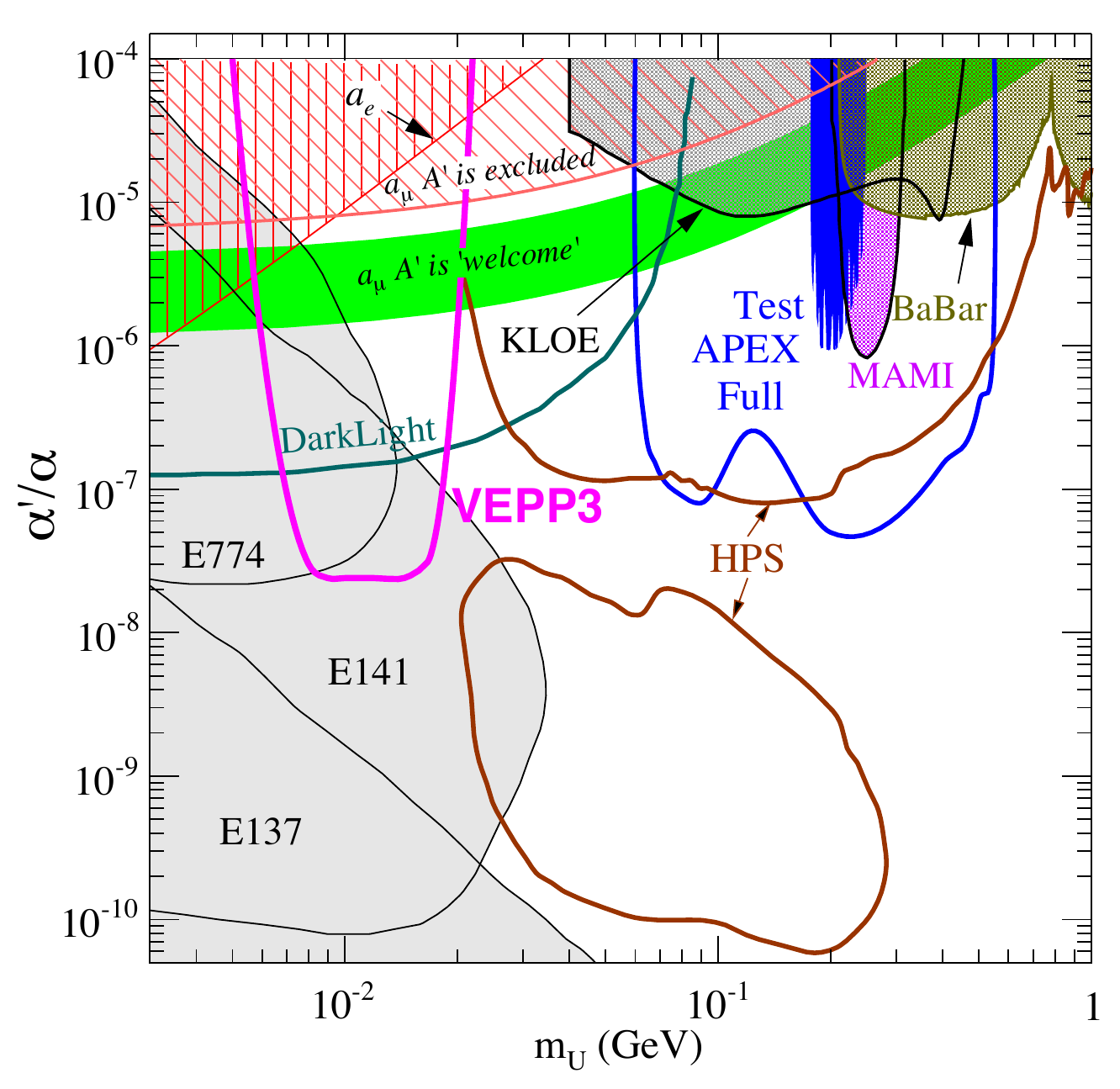}
\caption{Experimental reach of various (visible) $A'$ searches (taken from \Ref{Wojtsekhowski:2012zq}, where $A'$ is referred to as $U$).  Note that the green curve labeled ``DarkLight'' refers only to the visible $e^+ e^-$ search, and not to the invisible search discussed in this paper.  The VEPP-3 curve includes both the visible and invisible $A'$ search, since the search strategy is independent of the $A'$ decay mode.}
\label{fig:VEPP3Reach}
\end{center}
\end{figure}

The reach for the VEPP-3 experiment, along with the reach of several other experiments, is shown in \Fig{fig:VEPP3Reach}.  We emphasize that most of the curves are for the visible $A' \to e^+ e^-$ mode, while the VEPP-3 curve holds for any decay mode. The sharp threshold at $m_{A'} \approx 21.1 \MeV$ is clearly apparent (here the $A'$ is labeled $U$ on the horizontal axis).  VEPP-3 is mainly limited by the difficulty of attaining a high-intensity, high-energy positron beam, but is expected to have excellent performance up to the kinematic threshold for $A'$ production.

\begin{figure}[tp]
\begin{center}
\includegraphics[width=0.8\columnwidth]{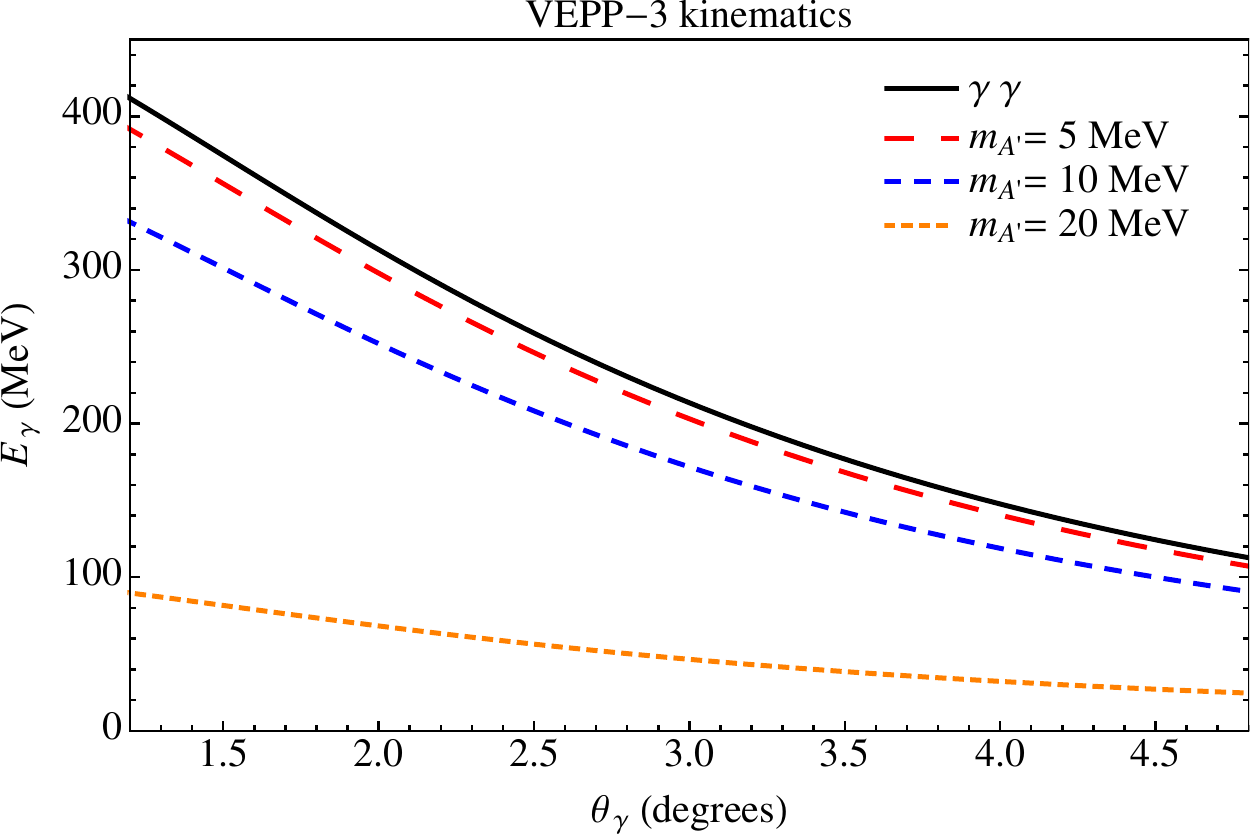}
\caption{Kinematic correlations between the measured photon energy and angle (lab frame) in the VEPP-3 setup for various $A'$ masses.  For large enough $m_{A'}$ there is a clean separation between the $e^+ e^- \to \gamma \gamma$ background and the $e^+ e^- \to \gamma A'$ signal.}
\label{fig:VEPP3Kinematics}
\end{center}
\end{figure}

The VEPP-3 setup has several interesting advantages. First, the process $e^+ e^- \to \gamma A'$ is a radiative return process, where the cross section continues to increase until just below the kinematic threshold for producing an $A'$.  This enhances the experimental reach despite the rather small center-of-mass energy due to the fixed target.  Second, the VEPP-3 signal is a $2 \to 2$  process, compared to the phase-space suppressed $2 \to 3$ signal for DarkLight.  This allows VEPP-3 to have a better reach than DarkLight despite having lower luminosity.  Finally, although the measured final state only consists of a single detected photon, the $2 \to 2$ kinematics is overconstrained.  As shown in \Fig{fig:VEPP3Kinematics}, the outgoing photon energy and scattering angle are perfectly correlated for a given $A'$ mass, meaning that measurement errors which caused $ep\gamma$ to fake a signal in the DarkLight setup will be less severe in the VEPP-3 setup.

As with DarkLight, VEPP-3 has a potentially huge background from the QED process where $A'$ is replaced by a photon, in this case $e^+ e^- \to \gamma \gamma$.  Since it has distinct kinematics from the signal, the VEPP-3 experiment is designed to eliminate this background using symmetric photon detectors which veto an event where two photons are detected in coincidence; this serves the same purpose as the DarkLight elastic veto and positive invariant mass cut.  That said, even with overconstrained kinematics, measurement errors with the quoted resolutions can still pose a problem for low masses because of the high $\gamma \gamma$ rate. The quoted energy and angular resolutions for VEPP-3 are $\sigma_E/E = 5\%$ for $100 \MeV < E < 450 \MeV$ and $\sigma_\theta = 0.1^\circ$, so looking at \Fig{fig:VEPP3Kinematics}, the $\gamma \gamma$ background and $m_{A'} = 5 \MeV$ signal are only separated by $1\sigma-1.5\sigma$.   This explains why the reach curve in \Fig{fig:VEPP3Reach} sharply degrades near $m_{A'} = 5 \MeV$.

The main challenge of the VEPP-3 setup is that the signal depends on detection of a single photon; any process which produces any number of final-state photons can potentially fake a signal. For example, $e^+ e^- \to \gamma \gamma \gamma$ can pose a significant background, as long as the kinematics are such that only a single photon ends up in the detector volume while the other two miss. Other backgrounds include ordinary bremsstrahlung off atomic electrons ($e^+ e^- \to e^+ e^- \gamma$) or protons ($e^+ p \to e^+ p \gamma$). These events can be partially suppressed with a detector looking for final-state positrons. As was anticipated in the VEPP-3 proposal, the positron bremsstrahlung cross section is the dominates through the allowed parameter space, even up to threshold. However, once a modest 10\% positron veto is implemented, the 3-photon process dominates. These events are irreducible in the same sense as $ep\gamma\gamma$ events for DarkLight, and limit the reach unless additional coincident photon detectors are added.

\subsection{Constraints from rare kaon decays}
\label{sec:Kaons}

As pointed out in \Ref{Fayet:2007ua} and calculated in \Ref{Pospelov:2008zw}, the rare kaon decay $K^+ \to \pi^+ + \textrm{inv.}$ can provide strong constraints on an $A'$ with couplings to quarks. The Standard Model contribution to this process $K^+ \to \pi^+ \nu \overline{\nu}$ is forbidden at tree-level by the absence of flavor-changing neutral currents (FCNCs), and further suppressed by the GIM mechanism, resulting in a branching ratio of $(8.4 \pm 1.0) \times 10^{-11}$. The measured branching ratio is Br$(K^+ \to \pi^+ \nu \overline{\nu}) = (1.7 \pm 1.1) \times 10^{-10}$ \cite{Beringer:1900zz}, in agreement with the Standard Model value, so this decay mode constrains any model of new physics which could contribute to it.   For an invisibly decaying $A'$, we must consider constraints on $K^+ \to \pi^+ A'$, which we do in an effective operator framework below the QCD confinement scale.

If the $A'$ couplings originate from kinetic mixing with the photon, then $K^+ \to \pi^+ A'$ process is still suppressed at tree-level since kinetic mixing leaves the flavor structure diagonal.  This logic excludes the operator $\partial_\mu K \pi A'^{\mu}$ in the effective theory because, replacing $A'^{\mu} \to A^{\mu}/\kappa$ where $\kappa = \sqrt{\alpha'/\alphaEM}$, we would get a flavor-changing tree-level coupling to the photon.   The lowest-dimension operator mediating $K^+ \to \pi^+ \gamma^*$ (note that the decay to an on-shell photon is forbidden by angular momentum conservation) appears at dimension-6:
\begin{equation}
\mathcal{L} \supset \frac{1}{\Lambda^2} \partial_\mu K \partial_\nu \pi F^{\mu \nu}.
\end{equation}
Via kinetic mixing, this generates the associated operator $(\kappa/\Lambda^2) \partial_\mu K \partial_\nu \pi F'^{\mu \nu}$ mediating $K^+ \to \pi^+ A'$. 

\begin{figure}[tp]
\begin{center}
\includegraphics[width=0.8\columnwidth]{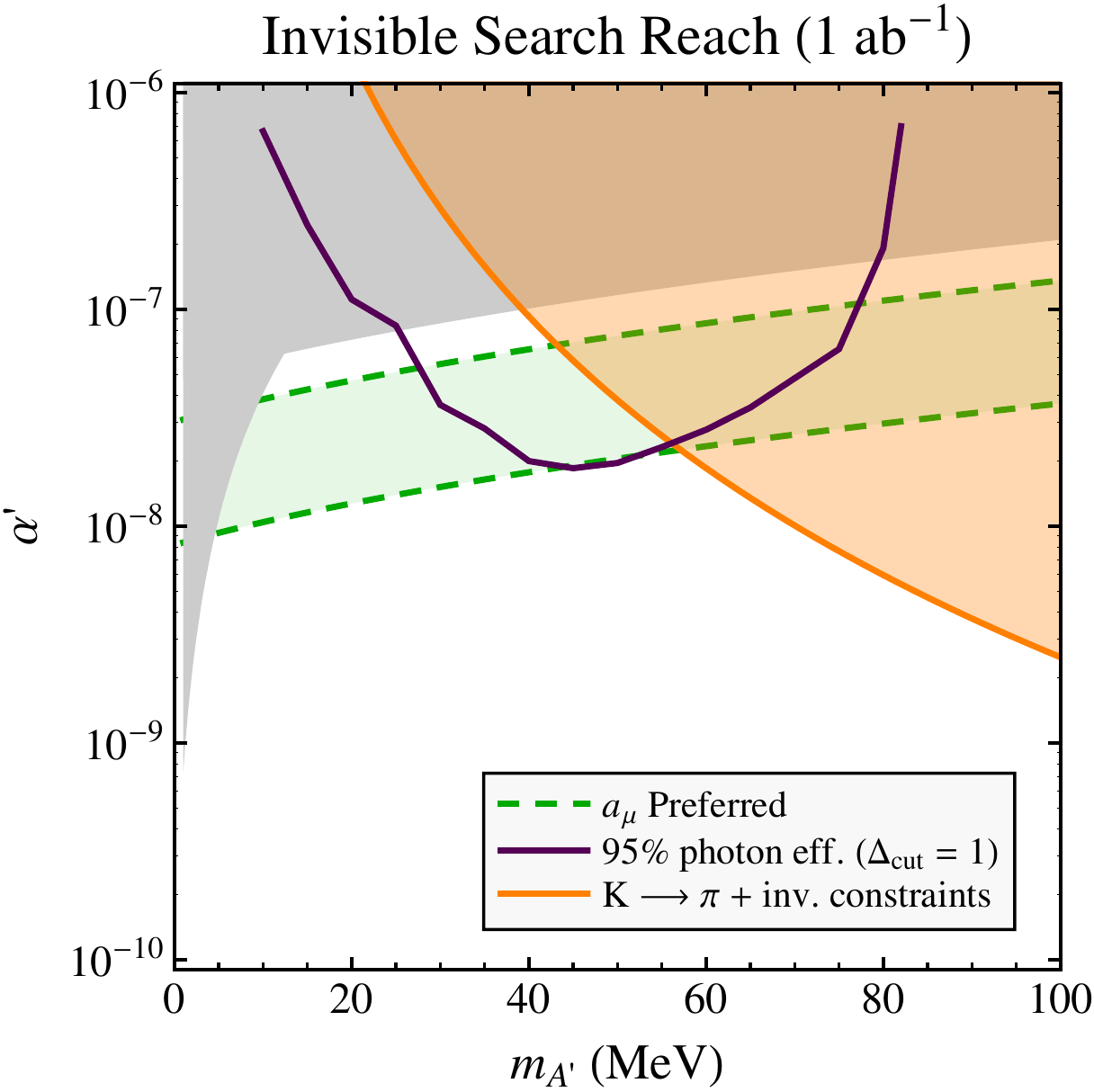}
\caption{Constraints on a kinetically-mixed $A'$ from the rare decay $K^+ \to \pi^+ + \textrm{inv.}$  The orange shaded region is excluded. This constraint would be removed if the $A'$ did not couple to quarks.}
\label{fig:KaonConstraints}
\end{center}
\end{figure}

Integrating by parts and using equations of motion, we have
\begin{align}
\mathcal{L} & \supset \frac{1}{\Lambda^2} \pi \partial_\mu K  \partial_{\nu}F^{\mu \nu} + \frac{\kappa}{\Lambda^2} \pi \partial_\mu K \partial_{\nu}F'^{\mu \nu} \nonumber\\
& \supset \frac{e}{\Lambda^2} \pi \partial_\mu K \overline{\psi}_e \gamma^\mu \psi_e + \frac{\kappa \, m_{A'}^2}{\Lambda^2}\pi \partial_\mu K A'^{\mu}.
\end{align}
The first operator mediates $K^+ \to \pi^+ e^+ e^-$; matching the observed branching ratio of $3 \times 10^{-7}$ \cite{Beringer:1900zz} for this process gives $\Lambda \approx 2~\textrm{TeV}$. Requiring the $K^+ \to \pi^+ + \textrm{inv.}$ branching ratio to stay within $2\sigma$ of the measured value after including the contribution from $K^+ \to \pi^+ A'$ gives the constraint shown in \Fig{fig:KaonConstraints}.  This result agrees with \Ref{Pospelov:2008zw}, which used matrix elements obtained from chiral perturbation theory \cite{D'Ambrosio:1998yj}.   

The above constraint is only valid if the $A'$ couplings arise from kinetic mixing.  But there are many model-dependent ways that this rare kaon decay constraint can be modified.  For example, if the $A'$ did not couple to quarks, this constraint would be absent entirely.  In the other extreme, the $A'$ could have tree-level flavor-violating couplings to the quarks, in which case the constraints would be considerably stronger.

\section{Conclusions}
\label{sec:Conclusions}

The DarkLight experiment is a unique probe of the luminosity frontier.  Despite the fact that elastic electron-proton scattering has been studied extensively since 1950 in an effort to understand the internal structure of the proton \cite{Rosenbluth:1950yq,Mcallister:1956ng}, we can use the increase in luminosity provided by the JLab FEL electron beam to tease out new physics which may be hiding at MeV-scale energies, unnoticed only because of its extremely weak coupling to the Standard Model. 

We have presented here an analysis strategy for DarkLight which could discover an invisibly-decaying vector boson $A'$ of mass $20-80\MeV$ in only 60 days of data taking, down to couplings of $\alpha' \approx 10^{-8}$. However, we have shown that such a discovery would only be possible with high photon detection efficiency, in order to suppress the large $ep\gamma \gamma$ background. It is remarkable that the ``sweet spot'' of this search lies in the region where discovery of an $A'$ could explain the $3\sigma$ anomaly in the muon magnetic moment.   A complementary experiment VEPP-3 is optimized to look for invisibly-decaying $A'$ bosons of mass $5-20\MeV$, and the combination of these two experiments can probe a large portion of parameter space, including the majority of the $(g-2)_\mu$ preferred region.

While our analysis has focused on the DarkLight setup in particular, we have introduced two new techniques which may be useful for other experiments at the luminosity frontier. We have  introduced an event selection criterion based on minimizing missing energy (in contrast with typical new physics cuts at the LHC which maximize missing energy).  This cut is useful in situations where the initial-state four-vectors are known with certainty, so that one can construct a missing invariant mass.  We have also described a method for mitigating the combinatorial background from pileup, by taking advantage of a peculiar feature of elastic kinematics which forces the mis-reconstructed invariant mass-squared to be negative. 

A discovery of a new massive vector boson $A'$ would open up the low-energy, high-luminosity frontier of particle physics. Since in many models the $A'$ mediates dark matter interactions, a discovery through its invisible decay mode could link dark matter physics with low-energy physics in a very interesting way.  Furthermore, DarkLight has the capability to detect both the visible and invisible decay modes, allowing in principle a measurement of the branching ratios to electrons/positrons and invisible final states.  Among experiments searching for a dark vector boson $A'$, this capability is unique to the DarkLight setup.   We eagerly await the construction of DarkLight and its contributions to physics beyond the Standard Model.

~\\

\textbf{Note added:} While this paper was being prepared, \Refs{Davoudiasl:2012ig, Endo:2012hp} appeared, using a new tenth-order calculation of the electron $g-2$ \cite{Aoyama:2012wj} to set more stringent bounds on $\alpha'$ in the low $A'$ mass region.

~\\

\begin{acknowledgments}
We thank Peter Fisher for many discussions on triggering, resolutions, and photon detection, and Narbe Kalantarians for preliminary photon efficiency estimates.  We would like to congratulate the DarkLight collaboration on a successful beam test in July 2012.   YK thanks Michael Schmitt and Tim Tait for early contributions to this project, and Bogdan Wojtsekhowski for helpful suggestions regarding backgrounds.  This work is supported by the U.S. Department of Energy (DOE) under cooperative research agreement DE-FG02-05ER-41360. YK is supported by an NSF Graduate Fellowship, and JT  is supported by the DOE Early Career research program DE-FG02-11ER-41741. JT acknowledges the hospitality of the Aspen Center for Physics, which is supported by the National Science Foundation Grant No.~PHY-1066293.
\end{acknowledgments}

\bibliography{InvisSearchBib}

\end{document}